\documentclass[tradiabstract]{aa}  
\newcommand{\bea}{\begin{eqnarray}}
\newcommand{\eea}{\end{eqnarray}}
\newcommand{\be}{\begin{equation}}
\newcommand{\ee}{\end{equation}}
\newcommand{\ben}{\begin{enumerate}}
\newcommand{\een}{\end{enumerate}}
\newcommand{\bi}{\begin{itemize}}
\newcommand{\ei}{\end{itemize}}
\newcommand{\bmi}[1]{\begin{minipage}{#1 cm}}
\newcommand{\emi}{\end{minipage}}

\newcommand{\rund}[1]{\left(#1\right)}
\newcommand{\vc}[1]{\mbox{\boldmath $#1$}}
\renewcommand{\d}{{\rm d}}
\newcommand{\wave}[1]{\left\{ #1 \right\}}
\newcommand{\eck}[1]{\left[ #1 \right]}
\newcommand{\ave}[1]{\left\langle #1 \right\rangle}
\newcommand{\vt}{\vartheta}
\newcommand{\vp}{\varphi}

\newcommand{\abs}[1]{| #1 |}

\def\arcminf {\hbox{$.\!\!^{\prime}$}}

\def\llabel#1{\label{sc:#1}}
\def\elabel#1{\label{eq:#1}}
\def\flabel#1{\label{fig:#1}}

{\catcode`\@=11
\gdef\SchlangeUnter#1#2{\lower2pt\vbox{\baselineskip 0pt \lineskip0pt
  \ialign{$\m@th#1\hfil##\hfil$\crcr#2\crcr\sim\crcr}}}
}

\newcommand{\beq}{\begin{eqnarray}}
\newcommand{\eeq}{\end{eqnarray}}

\usepackage{xcolor}

\usepackage{graphicx}
\usepackage{epsfig}
\usepackage{txfonts}
\usepackage{amsmath}
\usepackage{color}
\usepackage{natbib}
\usepackage{subfigure} 
\usepackage[switch, columnwise]{lineno}
\bibpunct{(}{)}{;}{a}{}{,}

\begin{document}

\def\tmin{{\vt_{\rm min}}}
\def\tmax{{\vt_{\rm max}}}

\title{Pure-mode correlation functions for cosmic shear and application to KiDS-1000}

   \author{Peter Schneider\inst{1}, Marika Asgari\inst{2,3}, Yasaman
     Najafi Jozani\inst{1}, Andrej Dvornik\inst{4}, Benjamin
     Giblin\inst{2},
     Joachim Harnois-D\'eraps\inst{5},
   Catherine Heymans\inst{2,4}, Hendrik Hildebrandt\inst{4}, Henk Hoekstra\inst{6}, Konrad Kuijken\inst{6}, HuanYuan Shan\inst{7,8}, Tilman Tröster\inst{2}, Angus H. Wright\inst{4}}

\offprints{Peter Schneider}
\institute{$^1$Argelander-Institut f\"ur Astronomie, 
  Universit\"at Bonn, Auf dem H\"ugel 71,
  D-53121 Bonn, Germany, \email{peter@astro.uni-bonn.de}\\
  $^2$Institute for Astronomy, University of Edinburgh, Royal
  Observatory, Blackford Hill, Edinburgh, EH9 3HJ, U.K.\\
  $^3$E. A. Milne Centre, University of Hull, Cottingham Road,
  Hull, HU6 7RX, UK\\
  $^4$Ruhr University Bochum, Faculty of Physics and Astronomy, Astronomical Institute (AIRUB), German Centre for Cosmological Lensing, 44780 Bochum, Germany\\
  $^5$School of Mathematics, Statistics and Physics, Newcastle University, Herschel Building, NE1 7RU, Newcastle-upon-Tyne, UK
  $^6$Leiden Observatory, Leiden University, P.O.Box 9513, 2300RA Leiden, The Netherlands\\
  $^{7}$Shanghai Astronomical Observatory (SHAO), Nandan Road 80,
Shanghai 200030, China \\
$^{8}$University of Chinese Academy of
Sciences, Beijing 100049, China
  }

\authorrunning{P. Schneider, M. Asgari, Y. Najafi et al.}


   \date{Received ; accepted }

   \abstract{One probe for systematic effects in gravitational lensing
     surveys is the presence of so-called B modes in the cosmic shear
     two-point correlation functions, $\xi_\pm(\vt)$, since lensing is
     expected to produce only E-mode shear. Furthermore, there exist
     ambiguous modes that cannot uniquely be assigned to either E-
     or B-mode shear. In this paper we derive explicit equations for
     the pure-mode shear correlation functions,
     $\xi_\pm^{\rm E/B}(\vt)$, and their ambiguous components,
     $\xi_\pm^{\rm amb}(\vt)$, that can be derived from the measured
     $\xi_\pm(\vt)$ on a finite angular interval,
     $\tmin\le\vt\le\tmax$, such that $\xi_\pm(\vt)$ can be decomposed
     uniquely into pure-mode functions as
     $\xi_+=\xi_+^{\rm E}+\xi_+^{\rm B}+\xi_+^{\rm amb}$ and
     $\xi_-=\xi_-^{\rm E}-\xi_-^{\rm B}+\xi_-^{\rm amb}$. The
     derivation is obtained by defining a new set of Complete
     Orthogonal Sets of E and B mode-separating Integrals (COSEBIs), for
     which explicit relations are obtained and which yields a smaller
     covariance between COSEBI modes. We derive the relation between
     $\xi_\pm^{\rm E/B/amb}$ and the underlying E- and B-mode power
     spectra.  The pure-mode correlation functions can provide a
     diagnostic of systematics in configuration space.  We then apply
     our results to Scinet LIght Cone Simulations (SLICS)
     and the Kilo-Degree Survey (KiDS-1000) cosmic shear data,
     calculate the new COSEBIs and the pure-mode correlation
     functions, as well as the corresponding covariances, and show
     that the new statistics fit equally well to the best fitting
     cosmological model as the previous KiDS-1000 analysis and
     recover the same level of (insignificant) B modes.  We also
     consider in some detail the ambiguous modes at the first- and
     second-order level, finding some surprising results. For example,
     the shear field of a point mass, when cut along a line through
     the center, cannot be ascribed uniquely to an E-mode shear and
     is thus ambiguous; additionally, the shear correlation functions resulting
     from a random ensemble of point masses, when measured over a
     finite angular range, correspond to an ambiguous mode.  }

   \maketitle
%

\section{\llabel{1} Introduction}

Statistical analysis of the weak distortions light bundles undergo
as they traverse the inhomogeneous Universe \citep{BSBV91, Kaiser92,
  Kaiser98} is believed to potentially be the most powerful empirical
probe for dark energy \citep{DETF, Peacock06}, provided systematic
effects can be controlled to a degree such that they are smaller than the
statistical error of large weak lensing surveys \citep[see, e.g.,][and
references therein]{Mandelbaum18}. A powerful demonstration of this
technique was provided by the Canada-France Hawaii Telescope Lensing Survey
\citep[CFHTLenS; see,
e.g.,][]{Heymans12,Heymans13,Erben13}, which revealed that the
amplitude of density fluctuations in the low-redshift Universe is
smaller than expected from the results obtained by measuring the
fluctuations of the cosmic microwave background (CMB). The current
generation of ground-based weak lensing surveys -- the Kilo Degree
Survey \citep[KiDS; e.g.,][]{Kuijken15, Kuijken19}, the Dark Energy
Survey \citep[DES; e.g.,][]{Sevilla-Noarbe21, Gatti21}, and the Hyper
SuprimeCam (HSC) Survey \citep[e.g.,][]{Aihara18} -- not only yield
impressive improvements over previous surveys in terms of survey area,
spectral coverage, and/or depth, but they have also led to a substantial
development of analysis tools regarding, for example, shear estimates and the
determination of the redshift distribution of source galaxies.
They have
also led to a consolidation of the tension regarding the level of
density fluctuations as measured by weak lensing and the CMB
(\citealt{Heymans21}, but see also \citealt{DESY3_3x2pt} for less discrepant
results; for a review on cosmological results from cosmic shear, see
\citealt{Kilbinger18}).

One of the tests for possible systematics in shear measurements
consists in the measurements of B-mode shear \citep{Crittenden02,
  SvWM02}. Gravitational lensing by the large-scale matter
distribution in the Universe is expected to yield some B-mode shear
due to lens-lens coupling, however with such a small amplitude that it
should remain undetectable even in all-sky surveys \citep{HHWS09,
  KrauseHirata10}.  The difference between shear and reduced shear
\citep{SchneiderSeitz95} affects the E-mode power spectrum
\citep[e.g.,][]{White05, Shapiro09, Deshpande20} but to the leading order
does not yield a B-mode contribution \citep{SvWM02}.  Other potential
sources of B-mode shear in data could be due to the clustering of
source galaxies \citep{SvWM02} or the inhomogeneous depth of
wide-field surveys \citep{Vale04,Heydenreich20}, but their amplitude
again is expected to be below the detection threshold. The expected
level of B modes from intrinsic alignments \citep[see, e.g.,][and
references therein]{Heymans06, Joachimi13/2, 2014MNRAS.437.1847G,
  Troxel15, Joachimi15, Hilbert17, Blazek19} is quite model dependent
and hence uncertain.  The most likely cause for any significant
B modes in shear data is thus the incomplete removal of systematic
effects, such as accounting for effects of the point-spread function.
For that reason, the significant detection of B modes in a shear
survey is considered a clear sign of remaining systematic
effects. We note that the opposite conclusion is not valid: the absence
of B modes does not imply that the data are systematics-free. For
example, a constant multiplicative bias would create no B modes but would
affect the E modes (see also \citealt{Kitching19} for more discussion on
this issue).

The most basic second-order shear statistics that can be derived from
survey data are the shear two-point correlation functions (2PCFs),
$\xi_\pm(\vt)$, since their estimates are unbiased by the presence of
gaps in the imaging data. Other second-order shear statistics can be
obtained as weighted integrals over $\xi_\pm(\vt)$. Of those, measures
that can separate E-mode shear from B-mode shear are of particular
interest. One such measure is the aperture mass dispersion, which was
introduced in \cite{Aperture98} and shown in \cite{SvWM02} to be
obtainable in terms of the shear correlation functions. However, as
pointed out by \cite{EBmix06}, the calculation of the aperture mass
dispersion requires knowledge of the shear correlation function
down to zero separation, which cannot be measured, for example due to the
overlapping images of galaxy pairs. The unavailability of $\xi_\pm$ at
very small angular scales then yields a bias in the aperture mass
statistics and a corresponding mixing of E and B modes. This issue
was addressed in \citet{Ring07}, where the general conditions for
E and B mode-separating second-order shear measures that can be obtained
from $\xi_\pm(\vt)$ on a finite interval of
$0<\tmin\le\vt\le\tmax<\infty$ were derived.

Based on this result, a Complete Orthogonal Set of E and B
mode-separating Integrals (COSEBIs) were defined in Schneider et
al.\,(\citeyear{SEK10}; hereafter SEK). The COSEBIs contain the
complete E and B mode-separable second-order shear information
obtainable from shear correlation functions on a finite angular
interval (see also \citealt{Becker13, BeckerRozo16} for a different
approach to decomposing the shear correlation functions into E-mode,
B-mode, and ambiguous mode statistics).  \cite{Asgari12} studied the
performance of COSEBIs on tomographic cosmic shear data, where shear
auto- and cross-correlation functions are measured from several source
galaxy populations with different redshift distributions. In these
papers it was demonstrated that the first few COSEBI components
contain essentially all the cosmological information, and hence they
serve as an efficient data compression method. Furthermore,
\cite{Asgari15} developed data compression further by defining
compressed COSEBIs (CCOSEBIs); they showed that even for tomographic
cosmic shear data the cosmologically relevant information is contained
in fewer than $\sim n_{\rm p}^2/2$ modes, where $n_{\rm p}$ is the
number of cosmological parameters.  In addition, COSEBIs are less
sensitive to density fluctuations on small spatial scales than the
shear correlation functions, for a given $\tmin$, and are therefore
less affected by ill-understood baryonic effects in structure
evolution \citep{Asgari20}.

In \cite{Asgari17}, COSEBIs and CCOSEBIs were applied to the CFHTLenS
cosmic shear data to probe for the presence of B-mode contributions
\citep[see also][for applications to other cosmic shear
data]{Asgari19a,Asgari19b}.  Using COSEBIs, \cite{Giblin21} and
\cite{Gatti21} showed that the most recent data sets from the KiDS
survey \citep[KiDS-1000; see][]{Kuijken19} and DES \citep[DES-Y3;
see][]{Sevilla-Noarbe21} show no indications of significant B-mode
shear. In addition, \cite{Asgari21} applied three different
second-order shear statistics to the KiDS-1000 shear data
\citep{Giblin21}, all of which yielded consistent results.

Whereas COSEBIs are extremely useful for extracting all E and B
mode-separable second-order information from a cosmic shear survey,
the interpretation of individual COSEBI modes is less
straightforward. Since they are not localized, neither in angular
space nor in Fourier space, a significant detection of B modes with
COSEBIs would be difficult to trace back to a given angular scale
\citep[see][for a thorough discussion on this point]{Asgari19a} and
thus to a possible origin of these B modes. A different approach for
separating modes consists in considering pure-mode shear correlation
functions, $\xi_{\pm{\rm E/B}}(\vt)$, which were first defined in
\cite{Crittenden02}; hereafter, we refer to them as CNPT correlation
functions, which corresponds to the initials of the authors of that
paper. However, estimating these CNPT correlation functions requires
the knowledge of the $\xi_\pm(\vt)$ for all angular scales. Due to the
lack of such measurements, previous applications of these CNPT
correlation functions (see, e.g., Hildebrandt et
al.\,\citeyear{Hildebrandt2017} and references therein) required an
extrapolation of $\xi_\pm$ to the smallest and largest angular scales,
or supplementing their measured values by theoretical predictions.

In this paper we derive a new set of pure-mode correlation functions
that we designate as $\xi_{\pm}^{\rm E/B}(\vt)$, which can be
calculated from the $\xi_\pm$ on a finite angular interval. These
pure-mode correlation functions can thus be obtained directly from the
data without extrapolation or modeling, and can hence be used to
study the angular dependence of any possible B-mode shear.

In order to derive $\xi_\pm^{\rm E/B}(\vt)$, we reconsider COSEBIs,
defining them with a slightly different orthogonality relation
relative to that used in SEK. In order to distinguish between these
two conventions, we denote the ones introduced by SEK as ``SEK
COSEBIs'' and the newly defined ones as ``dimensionless COSEBIs''
whenever the difference is relevant.  We show in
Sect.\,\ref{sc:EB-decomp} that for a given interval,
$\tmin\le\vt\le\tmax$, the shear correlation functions can be
decomposed into E modes, B modes, and ambiguous modes \citep[see
also][for a mode decomposition of CMB polarization data]{Bunn11}. The
ambiguous modes are contributions to the shear correlation functions
that cannot be uniquely ascribed to either E or B modes on a finite
separation interval but can be caused by either of them. In Appendix A
we consider in detail these ambiguous modes, both in terms of the
shear field and in terms of shear correlation functions and their
relation to the E- and B-mode power spectra. For example, we show
several examples of ambiguous shear correlation functions that can be
obtained from an E-mode power spectrum, a B-mode power spectrum, or a
mixture thereof. We note that ambiguous modes in the shear correlation
functions do occur because of the finite interval over which they are
measured. Indeed, formally setting $\tmin=0$ and $\tmax=\infty$, the
shear correlation functions can be uniquely decomposed into E and B
modes without ambiguous modes.

In Sect.\,\ref{sc:puremode} we define the pure-mode correlation
functions and derive closed-form expressions for them in terms of the
$\xi_\pm(\vt)$, discuss their general properties, show that the
COSEBIs can be obtained in term of the $\xi_{\pm}^{\rm E/B}$, compare
them to the CNPT correlation functions derived by \cite{Crittenden02},
to which they converge in the limit of $\tmin\to 0$ and
$\tmax\to\infty$, and obtain their relation to the E- and B-mode shear
power spectra. We then measure both the new dimensionless COSEBIs and
the pure-mode correlation functions for the tomographic data of
$\sim 1000$ square degrees of the Kilo Degree Survey \citep[KiDS-1000;
see][]{Asgari21,Heymans21} and compare them with the predictions from
the best fitting $\Lambda$ cold dark matter ($\Lambda$CDM) cosmology results of \cite{Asgari21}.
We also compare the performance of $\xi_{\pm}^{\rm E/B}$ with the CNPT
correlation functions using systematic-induced Scinet LIght Cone
Simulations \citep[SLICS;][]{SLICS18} following the methodology in
\cite{Asgari19a}.

We briefly summarize and discuss our main results in
Sect.\,\ref{sc:Summary}. Furthermore, in Appendix \ref{sc:newCOSEBI}
we present closed-form expressions for the new set of polynomial
weight functions for the COSEBIs that satisfy their modified
orthonormality relation that we employ in this paper, and we provide
an explicit code for calculating weight functions that are polynomial
in $\ln\vt$, yielding the logarithmic COSEBIs. We find that the
correlation matrix of the new COSEBIs has considerably smaller
off-diagonal elements, implying that the new set of COSEBIs yields
less mutual dependence than the previous one. Appendix
\ref{sc:subCOSEBI} explicitly shows that the COSEBIs related to a
subinterval of $\tmin$ and $\tmax$ can be obtained from those on the
full interval, and that the ambiguous modes within the subinterval do
not depend only on those of the full interval, but also on its
COSEBIs, implying that pure-mode information gets transferred to
ambiguous modes and is thus lost when considering subintervals.

\section{\llabel{EB-decomp}Decomposition into E and B modes}
In this paper we are mainly concerned with second-order shear
statistics, expressed in terms of shear correlation functions. We
assume throughout that these correlation functions are due to a
statistically homogeneous and isotropic shear field, so that the
correlation functions depend only on the modulus of the separation
vector. As we will show below, in this case the shear correlation
functions can be uniquely decomposed into E-, B-, and ambiguous modes,
irrespective of whether the observed shear is physical (e.g., obtained
from a potential) or partly caused by a systematic effect.  In
Appendix A we discuss the distinction between these three modes of a
shear field at the first-order level.

\subsection{General mode decomposition}
Throughout this paper we use the flat-sky approximation; for the
largest angular scale considered in practical examples later on (5
degrees), this is expected to be very accurate.  We denote by
$\xi_\pm(\vt)$ the 2PCFs of shear as a function of angular separation
$\vt$.  It was shown in \citet{Ring07} that an E- and B-mode
separation of second-order shear statistics is obtained from
the 2PCFs by
\begin{align}
\label{eq:EBmodes}
{\rm EE}&={1\over 2}\int_0^\infty \d\vt\;\vt\,\eck{T_+(\vt)\,\xi_+(\vt)
+T_-(\vt)\,\xi_-(\vt)} \;,\nonumber \\
{\rm BB}&={1\over 2}\int_0^\infty \d\vt\;\vt\,\eck{T_+(\vt)\,\xi_+(\vt)
          -T_-(\vt)\,\xi_-(\vt)} \;,
\end{align}
provided the two weight functions $T_\pm$ are related through
\be
\int_0^\infty\d\vt\,\vt\,{\rm J}_0(\ell\vt)\, T_+(\vt) 
=\int_0^\infty\d\vt\,\vt\,{\rm
  J}_4(\ell\vt)\, T_-(\vt)
\elabel{TpmJ04}
\ee
or, equivalently,
\begin{align}
T_+(\vt)&=T_-(\vt)+\int_\vt^\infty\d\theta\;\theta\,T_-(\theta)
\rund{{4\over\theta^2}-{12\vt^2\over \theta^4}}\;, \nonumber \\
T_-(\vt)&=T_+(\vt)+\int_0^\vt\d\theta\;\theta\,T_+(\theta)
\rund{ {4\over \vt^2}-{12\theta^2\over \vt^4}}\;,
\elabel{Tplusminus}
\end{align}
where ${\rm J}_i$ are Bessel functions of the first kind.  Then,
${\rm EE}$ and ${\rm BB}$ contain only E and B modes, respectively.
Furthermore, \citet{Ring07} showed that an E- and B-mode separation can be
obtained from the shear 2PCFs on a finite interval
$\tmin\le \vt\le \tmax$, provided that the function $T_+$ vanishes
outside this interval and satisfies the two conditions
\be
\int_{\tmin}^{\tmax}\d\vt\;\vt\,T_+(\vt)=0=
\int_{\tmin}^{\tmax}\d\vt\;\vt^3\,T_+(\vt) \;.
\elabel{conditions}
\ee
In this case, the function $T_-(\vt)$ as calculated from
Eq.\,(\ref{eq:Tplusminus}) also has finite support on the interval
$\tmin\le \vt\le \tmax$ and in addition satisfies the relations
\be
\int_{\tmin}^{\tmax}{\d\vt \over \vt}\;T_-(\vt)=0=
\int_{\tmin}^{\tmax}{\d\vt \over \vt^3}\;T_-(\vt) \;.
\elabel{conditions2}
\ee
The physical reason for conditions (\ref{eq:conditions}), as
explained in SEK, is that a constant shear, and a shear field linear
in angular position, cannot be uniquely ascribed to either E or
B modes; these ambiguous modes are therefore filtered out. In Appendix
A we also provide a physical interpretation of conditions
(\ref{eq:conditions2}). Furthermore, we note that in the hypothetical
case $\tmin=0$, conditions (\ref{eq:conditions2}) no longer
hold.\footnote{A specific example for ${\rm EE}$ and ${\rm BB}$ are
  the aperture dispersions, $M_{\rm ap}^2(\theta)$ and
  $M_\perp^2(\theta)$, considered in \cite{SvWM02}; for them,
  $\tmin=0$ and $\tmax=2\theta$.  In that case, the corresponding
  function $T_-(\vt)$ is nonnegative, and hence does not obey conditions (\ref{eq:conditions2}) -- see Fig.~1 in \cite{SvWM02}.}

\subsection{Complete sets of E and B modes on a finite interval} 
In SEK we constructed two complete orthogonal sets of functions
$T_{+n}(\vt)$ on the interval $\tmin\le\vt\le\tmax$, subject to the
constraints (\ref{eq:conditions}), one of them being polynomials in
$\vt$, the other being polynomials in $\ln\vt$.  Here, we consider
again complete sets of orthogonal functions on the same interval,
however with a slightly different metric. Specifically, we consider a
set of functions $T_{+n}(\vt)$, $n\ge 1$, that satisfy the
orthonormality relation
\be
\int_\tmin^\tmax \d\vt\;\vt\,T_{+m}(\vt)\,T_{+n}(\vt)
={B\over \bar\vt^2}\,\delta_{mn}\;
\elabel{ON-rel}
\ee
for all $m,n\ge 1$, and where each function $T_{+n}(\vt)$ satisfies
conditions (\ref{eq:conditions}). Here,
\be
\bar\vt={\tmin+\tmax\over 2}\;,\quad
B={\tmax-\tmin\over\tmax+\tmin}\;
\ee
are the mean angular scale within the interval and the relative width,
respectively. We note that $\tmin=(1-B)\bar\vt$, $\tmax=(1+B)\bar\vt$.
Explicit constructions of such function sets will be given in
Appendix\,\ref{sc:newCOSEBI}, where we choose $T_{+n}(\vt)$ to be
a polynomial in either $\vt$ or in $\ln\vt$, of order $n+1$.

For each of the $T_{+n}(\vt)$, we define the corresponding function
$T_{-n}(\vt)$ according to Eq.\,(\ref{eq:Tplusminus}). Interestingly,
the $T_{-n}$ also form an orthogonal set of functions on the interval
$\tmin\le\vt\le\tmax$, as we will demonstrate next. For this, we make
use of Eq.\,(\ref{eq:TpmJ04}) and the orthogonality relation of Bessel
functions to write
\be
T_{-n}(\vt)=\int_0^\infty\d\ell\;\ell\,{\rm J}_4(\ell\vt)
\int_\tmin^\tmax \d\theta\;\theta\,{\rm J}_0(\ell\theta)
\,T_{+n}(\theta)\;.
\elabel{transf} 
\ee
Carrying out the $\ell$ integration leads to the second of
Eqs.\,(\ref{eq:Tplusminus}), but for the present purpose, it is more
convenient to keep this presentation. We now show a convenient 
property.

{\it Lemma}: We consider two functions $F_+(\vt)$ and $F_+'(\vt)$,
defined for $\vt\ge 0$, and let $F_-(\vt)$ and $F_-'(\vt)$ be the
functions obtained from them by applying the transformation
\be
F_-(\vt)=\int_0^\infty\d\ell\;\ell\,{\rm J}_4(\ell\vt)
\int_0^\infty \d\theta\;\theta\,{\rm J}_0(\ell\theta)
\,F_+(\theta)\;.
\elabel{transff} 
\ee
Then,
\be
\int_0^\infty\d\vt\;\vt\,F_-(\vt)\,F_-'(\vt) 
=\int_0^\infty\d\vt\;\vt\,F_+(\vt)\,F_+'(\vt)  \;.
\elabel{Lemma}
\ee
The proof of the Lemma is rather straightforward: using transformation (\ref{eq:transff}), we obtain
\begin{align}
\int_0^\infty\d\vt&\;\vt\,F_-(\vt)\,F_-'(\vt) 
=\int_0^\infty\d\vt\;\vt
\int_0^\infty\d\ell\;\ell\,{\rm J}_4(\ell\vt)\nonumber \\
&\times \int_0^\infty \d\theta\;\theta\;{\rm J}_0(\ell\theta)
\,F_+(\theta) 
\int_0^\infty\d\ell'\;\ell'\,{\rm J}_4(\ell'\vt) \\
&\times
\int_0^\infty \d\theta'\;\theta'\,{\rm J}_0(\ell'\theta')
\,F_+'(\theta') \;.\nonumber
\end{align}
We now carry out the $\vt$ integration using
\be
\int_0^\infty\d\vt\;\vt\,{\rm J}_n(\ell\vt)\,{\rm J}_n(\ell'\vt)
={1\over\ell}\delta_{\rm D}(\ell-\ell')\;,
\elabel{J-ortho}
\ee
after which the $\ell'$ integration becomes trivial, yielding
\begin{align}
\int_0^\infty&\d\vt\;\vt\,F_-(\vt)\,F'_-(\vt)
=\int_0^\infty\d\ell\;\ell
\int_\tmin^\tmax \d\theta\;\theta\,{\rm J}_0(\ell\theta)
\,F_+(\theta)\nonumber \\
&\times\int_0^\infty \d\theta'\;\theta'\,{\rm J}_0(\ell\theta')
\,F_+'(\theta') =\int_0^\infty
\d\theta\;\theta\,F_+(\theta)\,F'_+(\theta) \;,
\end{align}
applying Eq.\,({\ref{eq:J-ortho}) again.  This completes the
proof.

We next apply the Lemma by letting $F_+=T_{+m}$, $F'_+=T_{+n}$; noting
that the $T_{+n}$ are zero outside the interval $\tmin\le\vt\le\tmax$,
we see from Eqs.\,(\ref{eq:transf}) and (\ref{eq:transff}) that 
$F_-=T_{-m}$, $F'_-=T_{-n}$. Therefore,
\be
\int_\tmin^\tmax\!\!\!\!\d\vt\;\vt\,T_{-m}(\vt)\,T_{-n}(\vt)
=\!\!\int_\tmin^\tmax\!\!\!\!\d\vt\;\vt\,T_{+m}(\vt)\,T_{+n}(\vt)
={B\over \bar\vt^2}\,\delta_{mn}\;.
 \elabel{Tminusortho}
\ee
Thus, the set of $T_{-n}(\vt)$ functions obeys the same orthogonality
relations as the $T_{+n}$.

In order to obtain a complete set of functions on the interval
$\tmin\le\vt\le\tmax$ irrespective of conditions
(\ref{eq:conditions}), we need to augment the set of the $T_{+n}$ by
two more functions that do not obey conditions
(\ref{eq:conditions}), which we call $T_{+a}(\vt)$ and
$T_{+b}(\vt)$. We choose them as
\be
T_{+a}(\vt)={1\over\sqrt{2}\bar\vt^2}\; ; \;
T_{+b}(\vt)={\sqrt{3}\over 2\sqrt{2} B\bar\vt^2}\eck{\rund{\vt\over\bar\vt}^2
-\rund{1+B^2}}\;.
\elabel{TaTb}
\ee
Both functions are normalized according to Eq.\,(\ref{eq:ON-rel}), and
they are mutually orthogonal. Furthermore, both of them are orthogonal
to all $T_{+n}(\vt)$ due to conditions (\ref{eq:conditions}). Thus,
the set of of functions $T_{+\mu}(\vt)$, $\mu=a, b, 1, 2, \dots$, form
a complete orthonormal set of functions on the interval
$\tmin\le\vt\le\tmax$.\footnote{From Eqs.\,(\ref{eq:conditions})
    and (\ref{eq:ON-rel}) it is obvious
  that $T_{+a}(\vt)$ and $T_{+b}(\vt)$ cannot be represented as a linear
  combination of the $T_{+n}(\vt)$; hence, the $T_{+n}(\vt)$ do not form a
  complete set of functions.}

We cannot use these two functions in Eq.\,(\ref{eq:Tplusminus}) to
obtain corresponding functions $T_{-a,b}$, since those would not have
finite support. Instead, we choose the two additional functions  
\begin{align}
T_{-a}(\vt)&={1-B^2\over \sqrt{2}\,\vt^2} \; , \nonumber\\
T_{-b}(\vt)&=\sqrt{3\over 8}{1-B^2\over B}
\eck{{1+B^2\over \vt^2}-{(1-B^2)^2 \bar\vt^2\over \vt^4}} \;,
\elabel{tminusdef}
\end{align}
which are orthogonal to all $T_{-n}$, according to
Eq.\,(\ref{eq:conditions2}), and obey the orthonormality relation
(\ref{eq:Tminusortho}). Thus, we now have two complete orthonormal
sets of functions on the interval $\tmin\le\vt\le\tmax$, the
$T_{+\mu}$, and the $T_{-\mu}$.

We now define the quantities $E_\mu$ and $B_\mu$ through
\begin{align}
E_\mu&={1\over 2}\int_\tmin^\tmax \d\vt\;\vt\,\eck{T_{+\mu}(\vt)\,\xi_+(\vt)
+T_{-\mu}(\vt)\,\xi_-(\vt)} \;,\nonumber \\
B_\mu&={1\over 2}\int_\tmin^\tmax \d\vt\;\vt\,\eck{T_{+\mu}(\vt)\,\xi_+(\vt)
-T_{-\mu}(\vt)\,\xi_-(\vt)} \;.
\label{eq:EBdef}
\end{align}
For $\mu=n$, with $n\ge 1$, these form the COSEBIs for the given set
of functions $T_{\pm n}$, such that $E_n$ depends only on E-mode
shear, and $B_n$ contains only B-mode shear. For $\mu=a,b$, $E_\mu$
and $B_\mu$ do not have an analogous interpretation. We note that the
orthonormality condition for the $T_{n\pm}$ used in this paper makes
the COSEBIs dimensionless, in contrast to those defined in SEK: From
Eq.\,(\ref{eq:ON-rel}), we see that dimension of the $T_{+n}$ is
$(\rm angle)^{-2}$, and since the $\xi_\pm$ are dimensionless, we see
from Eq.\,(\ref{eq:EBmodes}) that the ${\rm EE}$, ${\rm BB}$, and thus
the $E_\mu$ and $B_\mu$ are dimensionless.

Since the $T_{+\mu}$ and the $T_{-\mu}$ both form a complete
orthonormal set of functions, we can write the shear correlation
functions on the interval $\tmin\le\vt\le\tmax$ as a superposition,
\be
\xi_\pm(\vt)={\bar\vt^2\over B}\sum_\mu \tau_{\pm\mu}\,T_{\pm\mu}(\vt)\;.
\elabel{xiexp}
\ee
Taking the sum of Eqs.\,({\ref{eq:EBdef}), we find
\begin{align}
E_\mu+B_\mu&=\int_\tmin^\tmax \d\vt\;\vt\,T_{+\mu}(\vt)\,\xi_+(\vt) 
\nonumber\\
&={\bar\vt^2\over B}\sum_\nu \tau_{+\nu}
\int_\tmin^\tmax \d\vt\;\vt\,T_{+\mu}(\vt)\,T_{+\nu}(\vt)
=\tau_{+\mu}\;,
\elabel{EBplus}
\end{align}
where we inserted the expansion (\ref{eq:xiexp}) and made use of the
orthogonality relation (\ref{eq:ON-rel}). From the difference of
Eqs.\,({\ref{eq:EBdef}), we obtain in complete analogy
\be
E_\mu-B_\mu=\int_\tmin^\tmax \d\vt\;\vt\,T_{-\mu}(\vt)\,\xi_-(\vt) 
=\tau_{-\mu}\;,
\elabel{EBminus}
\ee
so that
\be
E_\mu={\tau_{+\mu}+\tau_{-\mu}\over 2}\; ;\qquad
B_\mu={\tau_{+\mu}-\tau_{-\mu}\over 2}\;.
\ee

\section{\llabel{puremode} Pure-mode correlation functions}
In this section we consider the pure-mode correlation functions;
more specifically, we show that the shear correlation functions
can be decomposed as
\begin{align}
\xi_+(\vt)&=\xi_+^{\rm E}(\vt)+\xi_+^{\rm B}(\vt)+\xi_+^{\rm amb}(\vt)
            \;,\nonumber \\
\xi_-(\vt)&=\xi_-^{\rm E}(\vt)-\xi_-^{\rm B}(\vt)+\xi_-^{\rm amb}(\vt)
            \;,
\elabel{decomposition}            
\end{align}
where the pure E- and B-mode correlation functions are defined in
terms of the COSEBIs,
\be
\xi_+^{\rm E}(\vt):={\bar\vt^2\over B}\sum_{n=1}^\infty
E_n\,T_{+n}(\vt) \; ;\quad
\xi_+^{\rm B}(\vt):={\bar\vt^2\over B}\sum_{n=1}^\infty
B_n\,T_{+n}(\vt) \;,
\elabel{xiEBplus}
\ee
\be
\xi_-^{\rm E}(\vt):={\bar\vt^2\over B}\sum_{n=1}^\infty
E_n\,T_{-n}(\vt) \; ;\quad
\xi_-^{\rm B}(\vt):={\bar\vt^2\over B}\sum_{n=1}^\infty
B_n\,T_{-n}(\vt) \;,
\elabel{xiEBminus}
\ee
and the $\xi_\pm^{\rm amb}$ correspond to ambiguous modes,
\be
\xi_\pm^{\rm amb}(\vt)={\bar\vt^2\over B}\sum_{\mu=a,b}\tau_{\pm\mu}\,
T_{\pm\mu}(\vt)\;.
\ee
In Sect.\,\ref{sc:GP} we consider general properties of these
pure-mode correlation functions. We express these as integrals over
the $\xi_\pm$ in Sect.\,\ref{sc:PMfromxi}; hence, in order to
calculate the pure-mode correlation functions, one does not need to
calculate the COSEBIs as intermediate step. Readers less interested in
the derivation of the results can find the final expressions for the
pure-mode correlation functions in Eqs.\,(\ref{eq:xiplusE},
\ref{eq:xiplusB}, \ref{eq:xiEminus}, \ref{eq:xiBminus}). In
Sect.\,\ref{sc:oldxis}, we compare our pure-mode correlation functions
to the CNPT correlation functions that were defined previously in
\cite{Crittenden02} and \cite{SvWM02}, but not confined to a finite
separation interval. Some consistency checks for the pure-mode
correlation functions are described in Sect.\,\ref{sc:Consistency},
and their relation to the power spectra is derived in
Sect.\,\ref{sc:ReltoP}.

\subsection{\label{sc:GP}General properties}
According to these definitions and constraints
(\ref{eq:conditions}) and (\ref{eq:conditions2}) that the basis
functions $T_{\pm n}$ have to satisfy, we find that
\be
\int_{\tmin}^{\tmax}\d\vt\;\vt\,\xi^{\rm E,B}_+(\vt)=0=
\int_{\tmin}^{\tmax}\d\vt\;\vt^3\,\xi^{\rm E,B}_+(\vt)\;,
\ee
\be
\int_{\tmin}^{\tmax}{\d\vt\over\vt}\,\xi^{\rm E,B}_-(\vt)=0=
\int_{\tmin}^{\tmax}{\d\vt\over\vt^3}\,\xi^{\rm E,B}_+(\vt)\;.
\ee
These relations show that the pure-mode correlation functions need to
have (at least) two roots in the interval $\tmin\le\vt\le\tmax$, and
hence their functional form can be expected to differ substantially
from $\xi_\pm(\vt)$. An example for this was shown in Fig.\,7 of SEK,
where an equivalent definition of the pure-mode correlation functions
was applied. Furthermore, since $T_{-n}(\tmin)=T_{+n}(\tmin)$ and 
$T_{-n}(\tmax)=T_{+n}(\tmax)$, we find that 
\be
\xi^{\rm E/B}_+(\tmin)=\xi^{\rm E/B}_-(\tmin)\; , \quad
\xi^{\rm E/B}_+(\tmax)=\xi^{\rm E/B}_-(\tmax)\;.
\elabel{boundcond}
\ee
As expected, the COSEBIs can be expressed in terms of the pure-mode
correlation functions, as we find from
Eqs.\,(\ref{eq:xiEBplus}, \ref{eq:xiEBminus}) by multiplying them with
$\vt T_{\pm m}(\vt)$ and integrating over $\vt$, making use of the
orthogonality relation (\ref{eq:Tminusortho}):
\begin{align}
  \int_\tmin^\tmax\d\vt\;\vt\,\xi_+^{\rm E}(\vt)\,T_{+m}(\vt)
  &=E_m=\int_\tmin^\tmax\d\vt\;\vt\,\xi_-^{\rm E}(\vt)\,T_{-m}(\vt)
    \nonumber \;, \\
  \int_\tmin^\tmax\d\vt\;\vt\,\xi_+^{\rm B}(\vt)\,T_{+m}(\vt)
  &=B_m=\int_\tmin^\tmax\d\vt\;\vt\,\xi_-^{\rm B}(\vt)\,T_{-m}(\vt)\;.
    \elabel{EBfromxiEB}
\end{align}
The foregoing equations allow us to find relations between
$\xi_+^{\rm E/B}(\vt)$ and $\xi_-^{\rm E/B}(\vt)$. We start with a
consistency relation, by using the definition (\ref{eq:xiEBplus}) and
replacing $E_n$ or $B_n$ by the first expression in (\ref{eq:EBfromxiEB}),
which yields
\begin{align}
  \xi_+^{\rm E/B}(\vt)&={\bar\vt^2\over B}\sum_{n=1}^\infty
  T_{+n}(\vt)\int\d\theta\;\theta\,\xi_+^{\rm E/B}(\theta)\,
                      T_{+n}(\theta)\nonumber \\
  &={\bar\vt^2\over B}\sum_\mu
  T_{+\mu}(\vt)\int\d\theta\;\theta\,\xi_+^{\rm E/B}(\theta)\,
                      T_{+\mu}(\theta) =\xi_+^{\rm E/B}(\vt)\;,
\end{align}
where in the second step we made use of the fact that $\xi_+^{\rm E/B}$
is orthogonal to $T_{+a}$ and $T_{+b}$, so that we could extend the
sum over all $\mu=a,b,1,2,\dots$. In the final step, we made use
of the completeness of the $T_{+\mu}$, which implies
\be
{\bar\vt^2\over B} \sum_\mu T_{+\mu}(\vt)\,T_{+\mu}(\theta)=
{1\over\theta}\,\delta_{\rm D}(\vt-\theta)\;.
\elabel{completeness}
\ee
Next we derive a relation between $\xi_+^{\rm E/B}$ and $\xi_-^{\rm E/B}$,
again using  Eqs.\,(\ref{eq:xiEBplus}) and (\ref{eq:EBfromxiEB}),
\begin{align}
  \xi_+^{\rm E/B}(\vt)&={\bar\vt^2\over B}\sum_{n=1}^\infty
  T_{+n}(\vt)\int\d\theta\;\theta\,\xi_-^{\rm E/B}(\theta)\,
                      T_{-n}(\theta) \;.
                      \elabel{xipasxim} 
\end{align}
We consider the sum
\begin{align}
  \sum_{n=1}^\infty&
                      T_{+n}(\vt)\,T_{-n}(\theta)
                            \nonumber \\
                      =&\sum_{n=1}^\infty
                         T_{+n}(\vt)\eck{T_{+n}(\theta)+ 
                      \int_\tmin^\theta\d\vp\;\vp\,T_{+n}(\vp)
                         \rund{{4\over\theta^2}-{12\vp^2\over\theta^4}}}\nonumber\\
                  =&\sum_\mu
                         T_{+\mu}(\vt)\eck{T_{+\mu}(\theta)+ 
                      \int_\tmin^\theta\d\vp\;\vp\,T_{+\mu}(\vp)
                         \rund{{4\over\theta^2}-{12\vp^2\over\theta^4}}}\nonumber\\
&-\sum_{\mu=a,b}  T_{+\mu}(\vt)\eck{T_{+\mu}(\theta)+ 
                      \int_\tmin^\theta\d\vp\;\vp\,T_{+\mu}(\vp)
                \rund{{4\over\theta^2}-{12\vp^2\over\theta^4}}}\;.\nonumber 
\end{align}
The sum over all $\mu$ can be carried out using the completeness
relation (\ref{eq:completeness}). For the sum over $\mu=a,b$, we can
calculate the term in the bracket, to find that for $\mu=a$ and
$\mu=b$, the result is of the form $a/\theta^2+b/\theta^4$, and hence
can be expressed in terms of the $T_{-a,b}(\theta)$. Thus, we find that
\begin{align}
  {\bar\vt^2\over B}&\sum_{n=1}^\infty
                      T_{+n}(\vt)\,T_{-n}(\theta)\nonumber \\
  &={\delta_{\rm D}(\vt-\theta)\over\vt}
    +{\rm H}(\theta-\vt)\rund{{4\over\theta^2}-{12\vt^2\over\theta^4}}\\
    &- X_a(\vt)T_{-a}(\theta)-X_b(\vt)T_{-b}(\theta)\;,\nonumber
\end{align}
where the $X_{a,b}(\vt)$ are quadratic functions of $\vt$ whose actual
form is of no relevance here. Inserting this result into
Eq.\,(\ref{eq:xipasxim}), and making use of the fact that
$\xi_-^{\rm E/B}$ is orthogonal to $T_{-a}$ and $T_{-b}$, we finally
find
\be
\xi_+^{\rm E/B}(\vt)=\xi_-^{\rm E/B}(\vt)+\int_\vt^\tmax
\d\theta\;\theta\,\xi_-^{\rm
  E/B}(\theta)\rund{{4\over\theta^2}-{12\vt^2\over\theta^4}} \;.
\elabel{xiEpasxiEm}
\ee
Thus, we obtain a relation between $\xi_+^{\rm E/B}$ and $\xi_-^{\rm E/B}$
that is very similar to the one between $\xi_+$ and $\xi_-$ in the
absence of B modes,
\be
\xi_+(\vt)=\xi_-(\vt)+\int_\vt^\infty
\d\theta\;\theta\,\xi_-(\theta)\rund{{4\over\theta^2}
  -{12\vt^2\over\theta^4}} \;,
\ee
except that the integral only extends to $\tmax$. We can see from
Eq.\,(\ref{eq:xiEpasxiEm}) that conditions (\ref{eq:boundcond})
are satisfied; for $\vt=\tmax$ this is trivial, and for $\vt=\tmin$,
it follows from the functional form of the integrand and the
orthogonality of $\xi_-^{\rm E/B}$ to the $T_{-a,b}$.

Using analogous steps, one can derive the inverse of the relation,
\begin{align}
  \xi_-^{\rm E/B}(\vt)&={\bar\vt^2\over
  B}\sum_{n=1}^\infty E_n\,T_{-n}(\vt)
                        \nonumber \\
  &=\xi_+^{\rm E/B}(\vt)+\int_\tmin^\vt \d\theta\;\theta\,\xi_+^{\rm
    E/B}(\theta)
    \rund{{4\over \vt^2}-{12\theta^2\over \vt^4}}\;,
    \label{eq:35}
\end{align}
again in close analogy to a corresponding relation between $\xi_+$ and
$\xi_-$ in the absence of B modes.
  
We would like to point out that the pure-mode correlation
functions $\xi_+^{\rm E/B}(\vt)$, $\xi_-^{\rm E/B}(\vt)$, and the
set of COSEBIs $E_n$ and $B_n$, respectively, contain exactly the
same information, as Eqs.\,(\ref{eq:xiEpasxiEm}, \ref{eq:35},
\ref{eq:xiEBplus}, \ref{eq:xiEBminus}) show that one of these
quantities can be derived from any of the other two. In practice,
this exact equivalence will apply only approximately, due to the
finite number of COSEBI modes and the finite binning of the
correlation functions; we demonstrate this issue in
Sect.\,\ref{sc:kids}.

\subsection{\label{sc:PMfromxi} Pure-mode correlation functions from
  $\xi_\pm$}
Obviously, we can calculate these pure-mode correlation functions from
the set of the $E_n$, $B_n$ that can be calculated from
Eqs.\,(\ref{eq:EBdef}). However, as we show here, they can also
be obtained without first calculating the (infinite) set of COSEBIs.
For that, we consider 
\begin{align}
\xi_+^{\rm E}(\vt)+\xi_+^{\rm B}(\vt) 
&={\bar\vt^2\over B}\rund{\sum_\mu
\tau_{+\mu}T_{+\mu}(\vt) -\tau_{+a}T_{+a}(\vt) 
-\tau_{+b} T_{+b}(\vt)}\nonumber\\
&=\xi_+(\vt) -\eck{\tau_{+a}U_{+a}(\vt) +\tau_{+b} U_{+b}(\vt) }
\;,
\elabel{xiEplusB}
\end{align}
where we made use of Eq.\,(\ref{eq:xiexp}) and defined for $\mu=a,b$
\be
U_{+\mu}(\vt)={\bar\vt^2\over B} T_{+\mu}(\vt)\;.
\ee
Thus, in order to calculate this sum, we only need the two
coefficients $\tau_{+a,b}$ that can be calculated from $\xi_+$ using
Eq.\,(\ref{eq:EBplus}). Similarly,
\begin{align}
\xi_+^{\rm E}(\vt)&-\xi_+^{\rm B}(\vt)
={\bar\vt^2\over B}\sum_{n=1}^\infty \tau_{-n}T_{+n}(\vt)\nonumber\\
&={\bar\vt^2\over B}\sum_{n=1}^\infty \tau_{-n}
\eck{T_{-n}(\vt)+\int_\vt^\tmax {\d\theta\over\theta}\,
T_{-n}(\theta)\rund{4-{12\vt^2\over \theta^2}}}\nonumber \\
&=\xi_-(\vt)+\int_\vt^\tmax {\d\theta\over\theta}\,
\xi_-(\theta)\rund{4-{12\vt^2\over \theta^2}}
\elabel{xiEBminusL}\\
&-{\bar\vt^2\over B}\sum_{\mu=a,b}\tau_{-\mu}
\eck{T_{-\mu}(\vt)+\int_\vt^\tmax {\d\theta\over\theta}\,
T_{-\mu}(\theta)\rund{4-{12\vt^2\over \theta^2}}} \;,\nonumber
\end{align}
where we used the relation (\ref{eq:Tplusminus}) between the $T_{+n}$
and $T_{-n}$ and the decomposition (\ref{eq:xiexp}). The expression in
the final bracket of Eq.\,(\ref{eq:xiEBminusL}) can be calculated,
using Eq.\,(\ref{eq:tminusdef}). For both $\mu=a,b$, the resulting
expressions are of the form $a + b \vt^2$, and thus can be written in
terms of the $U_{+\mu}$. We then find
\begin{align}
\xi_+^{\rm E}(\vt)-\xi_+^{\rm B}(\vt)
&=\xi_-(\vt)+\int_\vt^\tmax{\d\theta\over\theta}\,
\xi_-(\theta)\rund{4-{12\vt^2\over \theta^2}}\nonumber\\
&-\sum_{\mu=a,b}\tau_{-\mu}\,U_{-\mu}(\vt)\;,
\end{align}
where
\begin{align}
  U_{-a}(\vt)&={1-B\over\sqrt{2}B(1+B)^3}
               \eck{3\rund{\vt\over\bar\vt}^2-2(1+B)^2} \;,\nonumber\\
  U_{-b}(\vt)&={\sqrt{3}\over\sqrt{8}B^2}
               \eck{{(1-B)(1+4B+B^2)\over(1+B)^3}
               \rund{\vt\over\bar\vt}^2 - (1-B^2)}\;.
\end{align}
We then finally obtain for the pure mode correlation functions
\begin{align}
\xi_+^{\rm E}(\vt)&={1\over 2}\eck{\xi_+(\vt)+\xi_-(\vt)
+\int_\vt^\tmax {\d\theta\over\theta}\,
\xi_-(\theta)\rund{4-{12\vt^2\over \theta^2}}}\nonumber \\
&-{1\over 2} \eck{S_+(\vt)+S_-(\vt)}\;,
\elabel{xiplusE}\\
\xi_+^{\rm B}(\vt)&={1\over 2}\eck{\xi_+(\vt)-\xi_-(\vt)
-\int_\vt^\tmax{\d\theta\over\theta}\,
\xi_-(\theta)\rund{4-{12\vt^2\over \theta^2}}}\nonumber \\
&-{1\over 2} \eck{S_+(\vt)-S_-(\vt)} \;.
\elabel{xiplusB}
\end{align}
Here, we have defined
\begin{align}
S_+(\vt)&\equiv \sum_{\mu=a,b}\tau_{+\mu} U_{+\mu}(\vt)
=\int_\tmin^\tmax {\d\theta\;\theta\over \bar\vt^2}\, 
\xi_+(\theta)\,H_+(\vt,\theta) \;,
\elabel{Splus} \\
S_-(\vt) &\equiv \sum_{\mu=a,b}\tau_{-\mu} U_{-\mu}(\vt)
=\int_\tmin^\tmax {\d\theta\over \theta}\, \xi_-(\theta)\,H_-(\vt,\theta) \;,
\elabel{Sminus}
\end{align}
where
\begin{align}
H_+(\vt,\theta)&=\bar\vt^2\sum_{\mu=a,b}T_{+\mu}(\theta)\,U_{+\mu}(\vt)
\elabel{Hplus} \\
&={1\over 8B^3}\wave{4 B^2+3\eck{\rund{\vt\over \bar\vt}^2-1-B^2}
\eck{\rund{\theta\over \bar\vt}^2-1-B^2}} \;,\nonumber 
\end{align}
\begin{align}
H_-&(\vt,\theta)=\theta^2 
\sum_{\mu=a,b}T_{-\mu}(\theta)\,U_{-\mu}(\vt) 
\nonumber \\
&={(1-B)^2\over 8B^3}\Bigg\{3(1-B)^2\eck{(1+B)^4-(1+4B+B^2)\rund{\vt\over\bar\vt}^2}
\rund{\theta\over \bar\vt}^{-2} 
\nonumber \\
&+\eck{3(1+B)^2 \rund{\vt\over\bar\vt}^2
-\rund{3+6B+14B^2+6B^3+3B^4}} \Bigg\}\;,
\elabel{Hminus}
\end{align}
and where we made use of Eq.\,(\ref{eq:EBplus}) and the forgoing
expressions for the $U_{\pm\mu}$. We note that the functions
$S_\pm(\vt)$ are of the form $a+b\vt^2$, and thus correspond to a
shear correlation caused by ambiguous modes. Indeed, by adding the two
Eqs.\,(\ref{eq:xiplusE}) and (\ref{eq:xiplusB}), we obtain the first
of Eq.\,(\ref{eq:decomposition}), with
\be
\xi_+^{\rm amb}(\vt)=S_+(\vt) \;.  
\ee
It is important to realize that the final expressions for $S_\pm(\vt)$
are independent of the specific choice of the functions
$T_{\pm,a,b}$. It is easy to see that any ``rotation'' in the
two-dimensional subspace of functions that do not obey conditions
(\ref{eq:conditions}) or (\ref{eq:conditions2}), respectively, leaves
the forgoing expressions invariant.

We plot an example for the decomposition of the shear correlation
function $\xi_+$ into E modes and ambiguous modes in the upper panel
of Fig.\,\ref{fig:puremodexi}. For separations close to $\tmin$,
$\xi_+^{\rm E}(\vt)$ is close to $\xi_+(\vt)$; however, for larger
values of $\vt$, these two functions are markedly different, due to
the increasing amplitude of the ambiguous modes. As expected,
$\xi_+^{\rm E}(\vt)$ has two roots in the interval considered, whereas
$\xi_+(\vt)$ stays positive.

At first sight, one might wonder that the ambiguous correlation
function has a large amplitude. But what should be kept in mind is
that the information of this function is contained solely in two
numbers. In particular, as was shown in \cite{Asgari12}, they contain
little cosmological information even if assumed to be solely due to
E-mode shear.

\begin{figure}
  \includegraphics[width=\columnwidth]{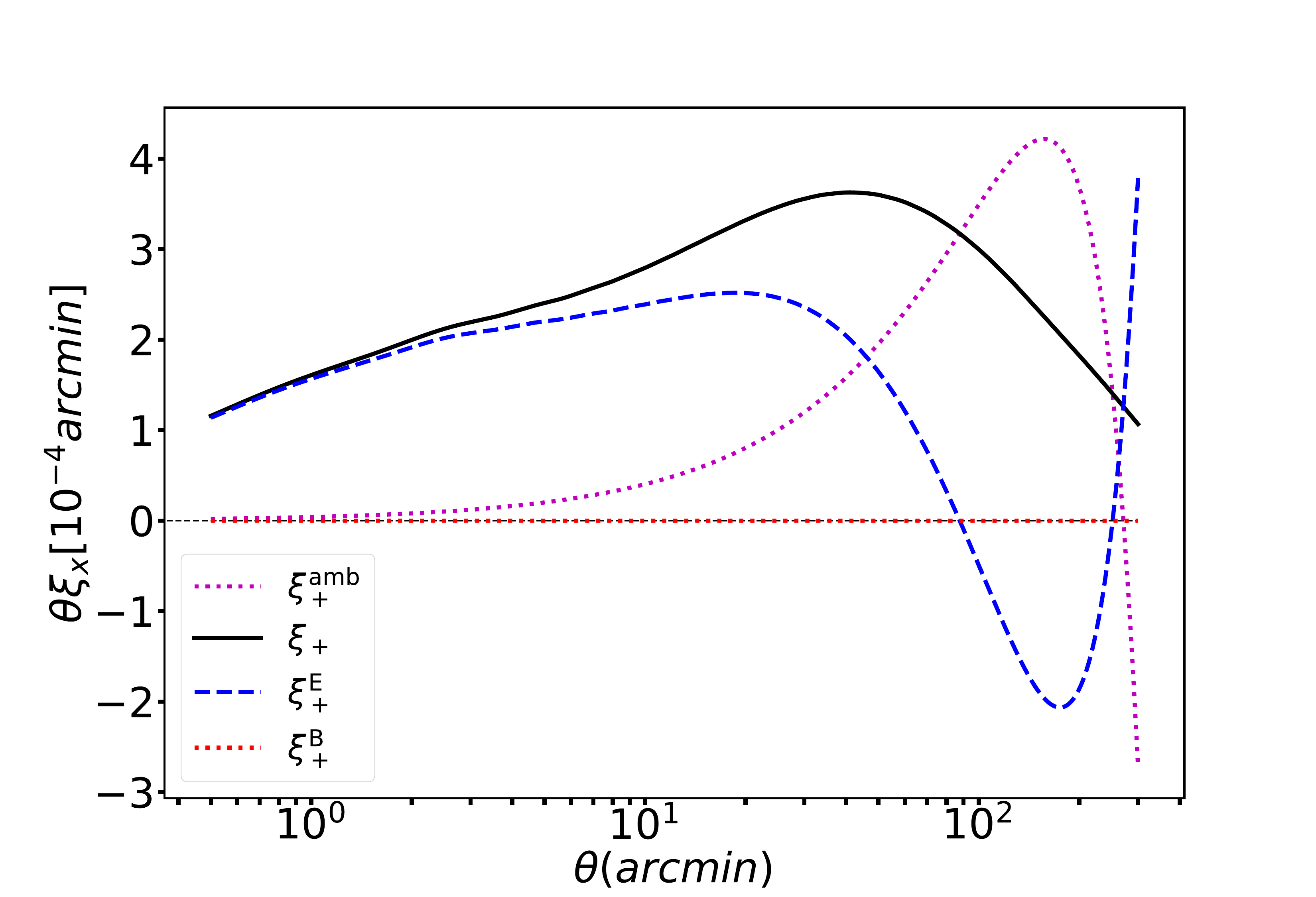}
  \includegraphics[width=\columnwidth]{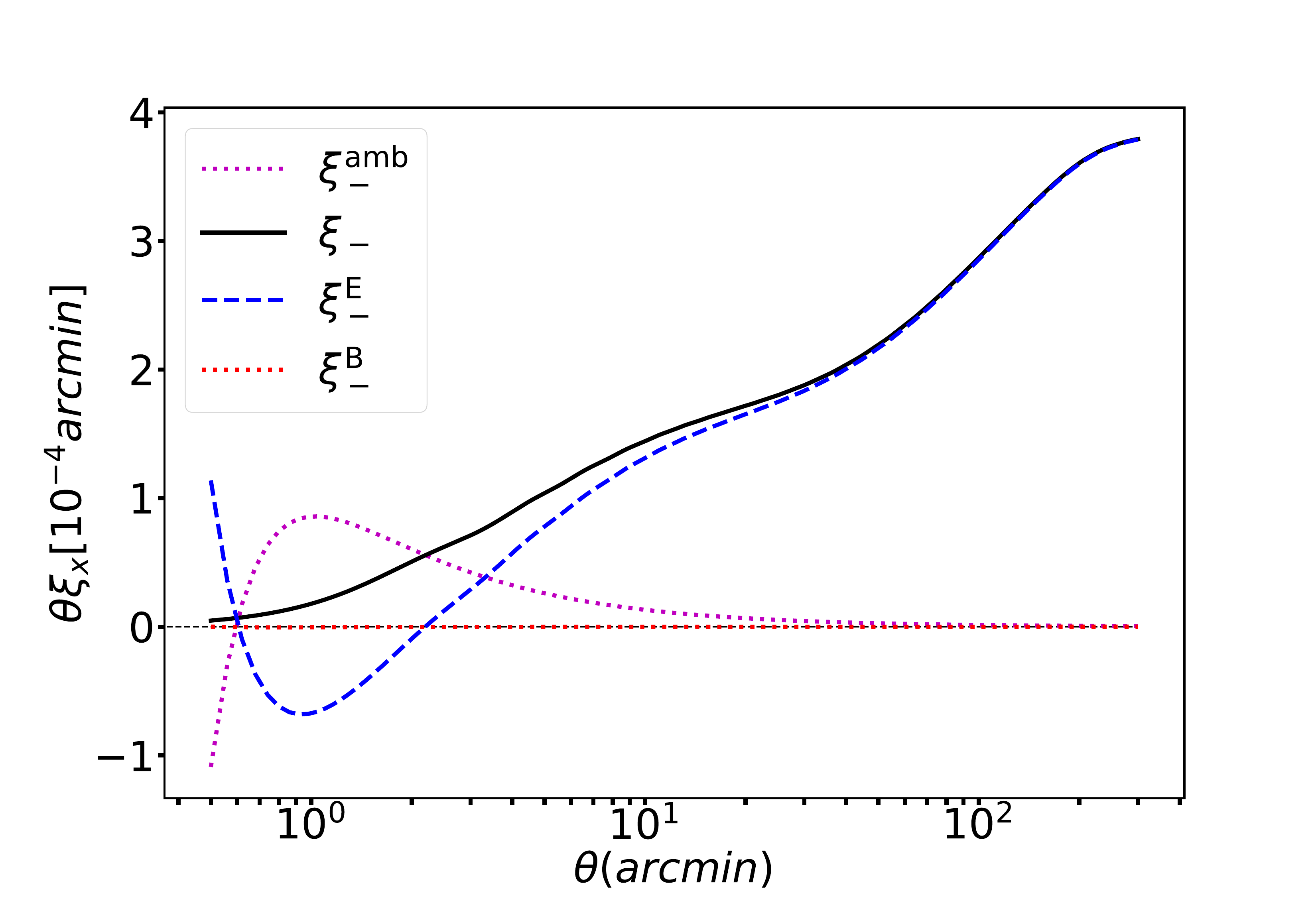}
  \caption{Decomposition of the shear correlation functions
    $\xi_+(\theta)$ (upper panel) and $\xi_-(\vt)$ (lower panel) into
    pure E modes (dashed blue curves) and ambiguous modes (dotted
    magenta curves). The latter are quadratic functions of $\theta$
    and $1/\theta$ for $\xi_+$ and $\xi_-$, respectively. We note that
    $\xi_+=\xi_+^{\rm E}+\xi_+^{\rm amb}$ due to the absence of
    B-mode shear assumed for this plot.  Here, we chose $\tmin=0\arcminf5$
    and $\tmax=300'$, and the correlation functions $\xi_\pm$ were
    calculated for a standard cosmological model fitted to the
    KiDS-1000 cosmic shear data (see Table \ref{tab:cosmoparams}). The
    source redshift distribution corresponds to the highest
    tomographic bin of the KiDS-1000 data.}
  \flabel{puremodexi}
\end{figure}

\begin{table}
\centering
\caption{Fiducial cosmological parameters.}
\label{tab:cosmoparams}
\begin{tabular}{ c  c  c  c  c  c  c}
\hline\\[-2.2ex]
$S_8$    &     $\Omega_{\rm m}$ & $\Omega_{\rm b}$ & $h$ & $n_{\rm s}$  &   $A_{\rm IA}$ & $A_{\rm bary}$ \\ \\[-2.0ex]
0.759  &   0.246 & 0.015 &     0.767 &     0.901 &     0.264 &     2.859 \\
\hline
\end{tabular}
\tablefoot{We employ a flat $\Lambda$CDM model with parameters fitted
  to the KiDS-1000 cosmic shear data \citep{Asgari21}. The structure
  growth parameter, $S_8=\sigma_8(\Omega_{\rm m}/0.3)^{0.5}$, and the
  amplitude of the intrinsic alignments of galaxies, $A_{\rm IA}$, are
  the only two parameters that KiDS-1000 cosmic shear data
  constrain. $\Omega_{\rm m}$ is the total matter density parameter,
  and $\Omega_{\rm b}$ represents the density parameter for baryonic
  matter. The spectral index of the primordial power spectrum is
  denoted as $n_{\rm s}$, while $h$ represents the dimensionless
  Hubble parameter.  We allow for baryonic feedback through
  $A_{\rm bary}$, which is equal to $3.13$ for a dark-matter-only
  scenario. Additionally, the sum of the neutrino masses is fixed to
  $0.06\,{\rm eV}$. }
\end{table}

Next, we turn to the ``$-$'' pure mode correlation functions. Using in
turn Eqs.\,(\ref{eq:xiEBminus}), (\ref{eq:xiexp}),
({\ref{eq:Tplusminus}), and (\ref{eq:Splus}), we find
\begin{align}
\xi_-^{\rm E}(\vt)&+\xi_-^{\rm B}(\vt)
={\bar\vt^2\over B}\sum_{n=1}^\infty \tau_{+n}T_{-n}(\vt)\nonumber\\
&={\bar\vt^2\over B}\sum_{n=1}^\infty \tau_{+n}\eck{T_{+n}(\vt)
+\int_\tmin^\vt {\d\theta\;\theta\over \vt^2}\,T_{+n}(\theta)\,\rund{4
   -{12\theta^2\over \vt^2}}} \nonumber\\
&=\xi_+(\vt)-S_+(\vt)+\int_\tmin^\vt
{\d\theta\;\theta\over \vt^2}\,\eck{\xi_+(\theta)-S_+(\theta)}
   \,\rund{4 -{12\theta^2\over \vt^2}} \nonumber\\
   &=\xi_+(\vt)+\int_\tmin^\vt
{\d\theta\;\theta\over \vt^2}\,\xi_+(\theta)
   \,\rund{4 -{12\theta^2\over \vt^2}} -V_+(\vt) \;,
\end{align}
where we have defined the function
\begin{align}
  V_+(\vt)&=S_+(\vt)+\int_\tmin^\vt
{\d\theta\;\theta\over \vt^2}\,S_+(\theta)
            \,\rund{4 -{12\theta^2\over \vt^2}}\nonumber \\
  &=\int_\tmin^\tmax {\d\theta\;\theta\over \bar\vt^2}\,\xi_+(\theta)
    \,K_+(\vt,\theta)\;,
\end{align}
and by using the definition (\ref{eq:Splus}) for $S_+$, we obtain for
the kernel $K_+$ the following expression:
\begin{align}
K_+&(\vt,\theta)=H_+(\vt,\theta)+\int_\tmin^\vt
{\d\vp\;\vp\over \vt^2}\,H_+(\vp,\theta)
\rund{4-{12 \vp^2\over \vt^2}} \nonumber\\
&= {(1-B)^2\over 8 B^3}
\Bigg\{ 3 (1-B)^2\rund{\vt\over \bar\vt}^{-4}
\eck{(1+B)^4-(1+4B+B^2)\rund{\theta\over\bar\vt}^2}\nonumber \\
&+\rund{\vt\over \bar\vt}^{-2}
\eck{3(1+B)^2\rund{\theta\over\bar\vt}^2-(3+6B+14 B^2+6B^3+3B^4)}
\Bigg\} \nonumber \\
&=\rund{\bar\vt\over\vt}^2 H_-(\theta,\vt) \; .
\end{align}
For the difference of the two ``$-$'' correlation functions we obtain
\be
\xi_-^{\rm E}(\vt)-\xi_-^{\rm B}(\vt)
={\bar\vt^2\over B}\sum_{n=1}^\infty \tau_{-n}T_{-n}(\vt)
  =\xi_-(\vt)-V_-(\vt) \;,
    \elabel{xiEminusB}
\ee
where
\be
V_-(\vt)={\bar\vt^2\over  B}
         \sum_{\mu=a,b}\tau_{-\mu}T_{-\mu}(\vt)
=\int_\tmin^\tmax{\d\theta\;\theta\over\bar\vt^2}\,
\xi_-(\theta)\,K_-(\vt,\theta)\;,
\ee
with the kernel function
\begin{align}
K_-(\vt,\theta)&={\bar\vt^4\over B}\sum_{\mu=a,b}
T_{-\mu}(\vt) T_{-\mu}(\theta)\nonumber \\
&={\bar\vt^4(1-B^2)^2\over B\vt^2\theta^2}
\Bigg\{{1\over 2}+{3\over 8B^2}\eck{1+B^2-(1-B^2)^2\rund{\bar\vt\over \vt}^2}\nonumber \\
&\quad\times\eck{1+B^2-(1-B^2)^2\rund{\bar\vt\over \theta}^2} \Bigg\} \\
&={(1-B^2)^2 \bar\vt^4\over \vt^2 \theta^2}
H_+\rund{ {[1-B^2] \bar\vt^2\over\vt}, {[1-B^2]
    \bar\vt^2\over\theta}}
\;.\nonumber
\end{align}
Therefore,
\begin{align}
\xi_-^{\rm E}(\vt)&={1\over 2}\eck{\xi_+(\vt)+\xi_-(\vt)
+\int_\tmin^\vt
{\d\theta\;\theta\over \vt^2}\,\xi_+(\theta)
\rund{4-{12\theta^2\over \vt^2}}} \nonumber \\
&-{1\over 2}\eck{V_+(\vt)+V_-(\vt)}\;,
\elabel{xiEminus}
\end{align}
\begin{align}
\xi_-^{\rm B}(\vt)&={1\over 2}\eck{\xi_+(\vt)-\xi_-(\vt)
+\int_\tmin^\vt
{\d\theta\;\theta\over \vt^2}\,\xi_+(\theta)
\rund{4-{12\theta^2\over \vt^2}}} \nonumber\\
&-{1\over 2}\eck{V_+(\vt)-V_-(\vt)}\;.
\elabel{xiBminus}
\end{align}
The functions $V_\pm(\vt)$ are of the form $a\vt^{-2}+b\vt^{-4}$, and
therefore correspond to shear correlations due to ambiguous
modes. These are subtracted from the rest of the expression to yield
pure E- and B-mode correlation functions. By subtracting
Eq.\,(\ref{eq:xiBminus}) from Eq.\,(\ref{eq:xiEminus}), we obtain the
second of Eqs.\,({\ref{eq:decomposition}), with
\be
\xi_-^{\rm amb}(\vt)=V_-(\vt) \;.
\ee
An example for the decomposition of $\xi_-$ into E- and ambiguous
modes is shown in the lower panel of Fig.\,\ref{fig:puremodexi}. For
large values of $\vt$, $\xi_-^{\rm E}$ differs only little from
$\xi_-$, but their difference increases for smaller $\vt$. In
particular, $\xi_-^{\rm E}$ has two roots in the interval
$\vt\in[\tmin,\tmax]$.

We point out that pure-mode correlation functions equivalent to the
foregoing ones were already defined in SEK. However, their expressions
in terms of $\xi_\pm$ in SEK were considerably more complicated than
the present ones, and therefore, they have not been applied to any
data, as far as we know. Our choice of the orthonormality relation,
which differs from the one in SEK, allowed us to obtain far more
explicit expressions for the pure-mode shear correlation functions,
and they are easily applicable to a set of measured $\xi_\pm$, as we
show in Sect.\,\ref{sc:kids}.

For completeness, we also note that in the case $\tmin=0$,
$\xi_-^{\rm amb}(\vt)\equiv 0$. In that case, $B=1$, and thus
$T_{-a}(\vt)\equiv 0\equiv T_{-b}(\vt)$.

\subsection{\label{sc:oldxis} Comparison with ``old'' pure-mode shear correlation functions}
\subsubsection{General considerations}
Previously, the CNPT correlation functions that were defined by
\cite{Crittenden02} and \cite{SvWM02} also yield a mode separation;
they are given in terms of the E- and B-mode convergence power spectra
$P_{\rm E,B}(\ell)$ through
\begin{align}
\xi_{\rm E,B+}^{\rm CNPT}(\vt)&=\int_0^\infty{\d\ell\;\ell\over 2\pi}\,P_{\rm
  E,B}(\ell)\,{\rm J}_0(\ell\vt) \;,\nonumber \\
\xi_{\rm E,B-}^{\rm CNPT}(\vt)&=\int_0^\infty{\d\ell\;\ell\over 2\pi}\,P_{\rm
  E,B}(\ell)\,{\rm J}_4(\ell\vt) \;.
\end{align}
These functions can be expressed solely in terms of the shear
correlation functions,
\begin{align}
\xi_{\rm E+}^{\rm CNPT}(\vt)&={1\over
  2}\eck{\xi_+(\vt)+\xi_-(\vt)+\int_\vt^\infty{\d\theta\over \theta}\;
\xi_-(\theta)\rund{4-{12\vt^2\over \theta^2}}} \;,\nonumber \\
\xi_{\rm E-}^{\rm CNPT}(\vt)&={1\over
  2}\eck{\xi_+(\vt)+\xi_-(\vt)+\int_0^\vt {\d\theta\;\theta\over \vt^2}\;
\xi_+(\theta)\rund{4-{12\theta^2\over \vt^2}}} \;,\nonumber \\
\xi_{\rm B+}^{\rm CNPT}(\vt)&={1\over
  2}\eck{\xi_+(\vt)-\xi_-(\vt)-\int_\vt^\infty{\d\theta\over \theta}\;
\xi_-(\theta)\rund{4-{12\vt^2\over \theta^2}}} \;, \elabel{xiEBold}\\
\xi_{\rm B-}^{\rm CNPT}(\vt)&={1\over
  2}\eck{\xi_+(\vt)-\xi_-(\vt)+\int_0^\vt {\d\theta\;\theta\over \vt^2}\;
\xi_+(\theta)\rund{4-{12\theta^2\over \vt^2}}} \;.\nonumber
\end{align}
These can now be compared to the pure-mode correlation functions on a
finite interval. We see that the functional form differs in two
respects. First, the integrals over the correlation functions
$\xi_\pm$ only extend over the finite interval for
$\xi^{\rm E,B}_\pm$, whereas they extend to either $0$ or $\infty$ for
$\xi_{\rm E,B\pm}^{\rm CNPT}$. Second, in the $\xi^{\rm E,B}_\pm$ a
term that corresponds to the ambiguous modes is subtracted.

Another way to see the difference between the CNPT and the pure-mode 
correlation functions is by noting that
\be
  \xi_+(\vt)=\xi_{\rm E+}^{\rm CNPT}(\vt)+\xi_{\rm B+}^{\rm CNPT}(\vt) \;;
  \quad
  \xi_-(\vt)=\xi_{\rm E-}^{\rm CNPT}(\vt)-\xi_{\rm B-}^{\rm CNPT}(\vt) \;,
\ee
whereas the decomposition into the pure-mode correlation functions is
given by Eq.\,(\ref{eq:decomposition}).

The $\xi_{\rm E,B\pm}^{\rm CNPT}$ are unobservable as they require a
measurement of $\xi_\pm$ either down to zero separation or up to
infinite separation; neither is possible. We note that the
$\xi_{\rm E,B\pm}^{\rm CNPT}$ do not account for ambiguous modes,
since for an infinite field, there are no ambiguous modes: a constant
shear on an infinite field would violate the assumption of statistical
isotropy of the random field (whereas on a collection of finite
fields, the constant shear can have random magnitude and orientations
for each field), and a linear shear field on an infinite field in
addition would diverge (see the discussion in Appendix
\ref{sc:Ambig}). The ambiguous mode $\xi_+^{\rm amb}$ is due to the
lack of information on $\xi_\pm$ for scales $\vt>\tmax$, whereas the
$\xi_-^{\rm amb}$ is rooted in the missing information from scales
$\vt<\tmin$.

We can check that the pure-mode shear correlation functions tend
toward the CNPT correlation functions in the limit $\tmin\to 0$ or
$\tmax\to\infty$. We consider first the ``$+$'' modes and let
$\tmax\to\infty$, which also implies $\bar\vt\to \infty$ and $B\to 1$
such that $(1-B)=\tmin/\bar\vt$. In this limit, the function
$H_+(\vt,\theta)$ tends to a constant, and $S_+(\vt)\to
0$. Furthermore, $H_-(\vt,\theta)\to 0$, due to the factors $(1-B)^2$
in Eq.\,(\ref{eq:Hminus}); correspondingly, $S_-(\vt)\to 0$. Thus, in
this limit, expressions (\ref{eq:xiplusE}) and (\ref{eq:xiplusB})
for $\xi_+^{\rm E/B}(\vt)$ converge to the corresponding ones in
Eq.\,(\ref{eq:xiEBold}). For the ``$-$'' modes, we consider
$\tmin\to 0$, implying $B\to 1$. That means that
$K_\pm(\vt,\theta)\to 0$, and thus $V_\pm(\vt)\to 0$. Hence, we see
that expressions (\ref{eq:xiEminus}) and (\ref{eq:xiBminus}) for
$\xi_-^{\rm E/B}(\vt)$ converge to the corresponding ones in
Eq.\,(\ref{eq:xiEBold}).

\subsubsection{Comparison using SLICS}
\cite{Asgari19a} modeled multiple data systematics that may exist in
cosmic shear data.  They applied these systematics to mock data from
SLICS N-body simulations (see their Sect. 6 for details). Ten lines-of-sight were chosen and the measurements were applied to shape-noise-free mock data.  Aside from the SEK COSEBIs they measured
$\xi_{\rm E/B+}^{\rm CNPT}$ from these simulations. Here we compare
the pure mode correlation functions with their measurements.

Figure\thinspace\ref{fig:slicsfid} compares the measured signal for
both the pure-mode and the CNPT correlation functions.  The results
are shown for the mean of ten lines-of-sight. Here the mock data are
free of systematic effects. The measurements are made for 50
logarithmic bins in $\theta$. As can be seen, these two sets of
correlation functions match at small separations, while they differ on
larger scales; this is because ambiguous modes are not removed from
$\xi_{\rm E/B+}^{\rm CNPT}$. In addition, a theoretical prediction for
$\xi_-$ is used beyond $\theta=300'$, to calculate the integrals in
Eq.\,(\ref{eq:xiEBold})}. In particular, we can see that the pure mode
$\xi_+^{\rm B}$ closely recovers the zero B-mode prediction, in
contrast to $\xi_{\rm B+}^{\rm CNPT}$.

\begin{figure}
  \includegraphics[width=\columnwidth]{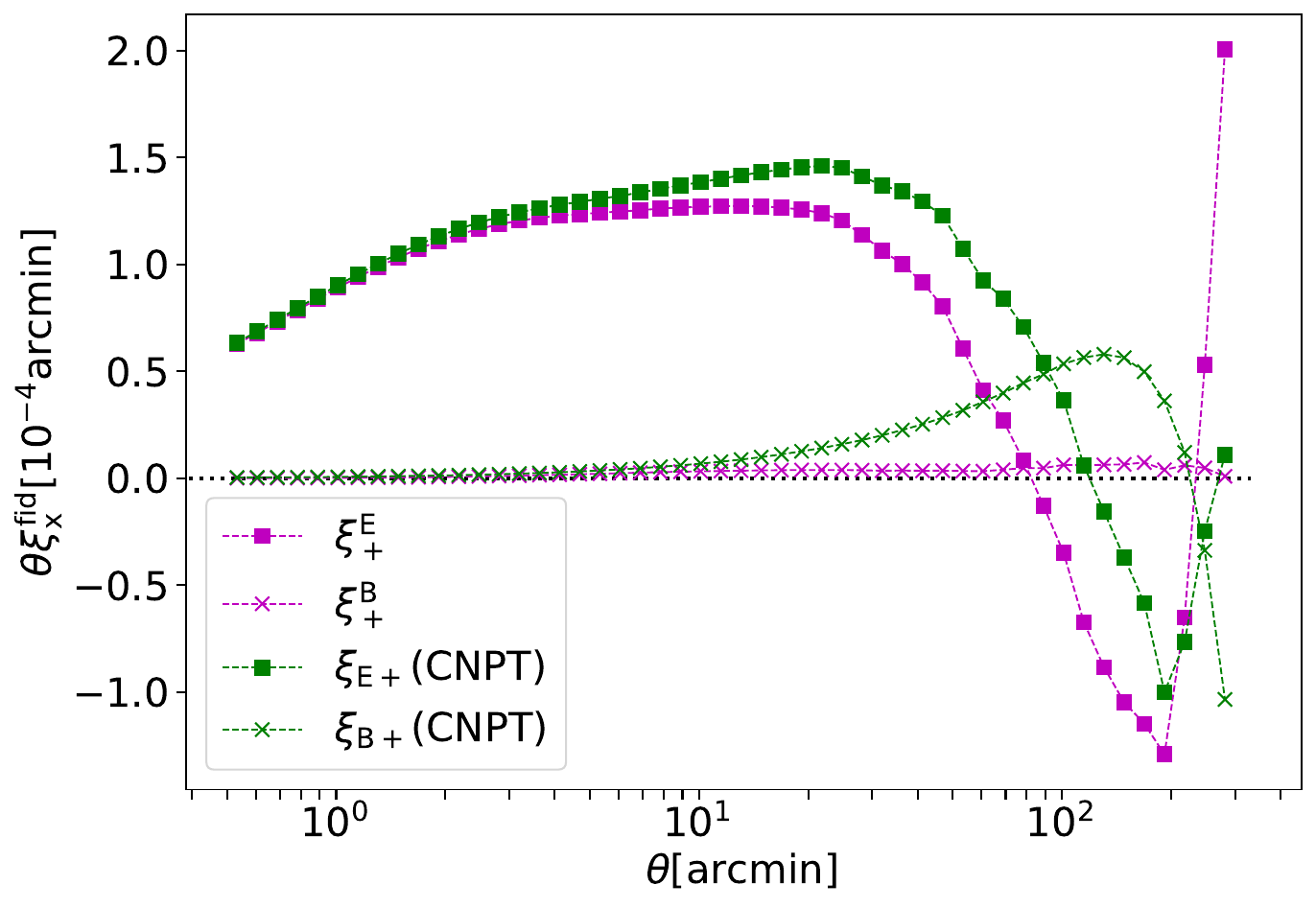}
  \caption{Measured E- and B-mode correlation functions from SLICS
    simulations. Both E modes (squares) and B modes (crosses) are
    averaged over ten shape-noise-free lines-of-sight. The pure-mode
    correlation functions (magenta) are insensitive to information
    outside of the defined angular separation range,
    $[0\arcminf5,300']$. The CNPT correlation functions (green) include
    ambiguous modes and information from outside of the measured
    range. }
  \flabel{slicsfid}
\end{figure}

\begin{figure*}[tbh]
  \includegraphics[width=0.99\columnwidth]{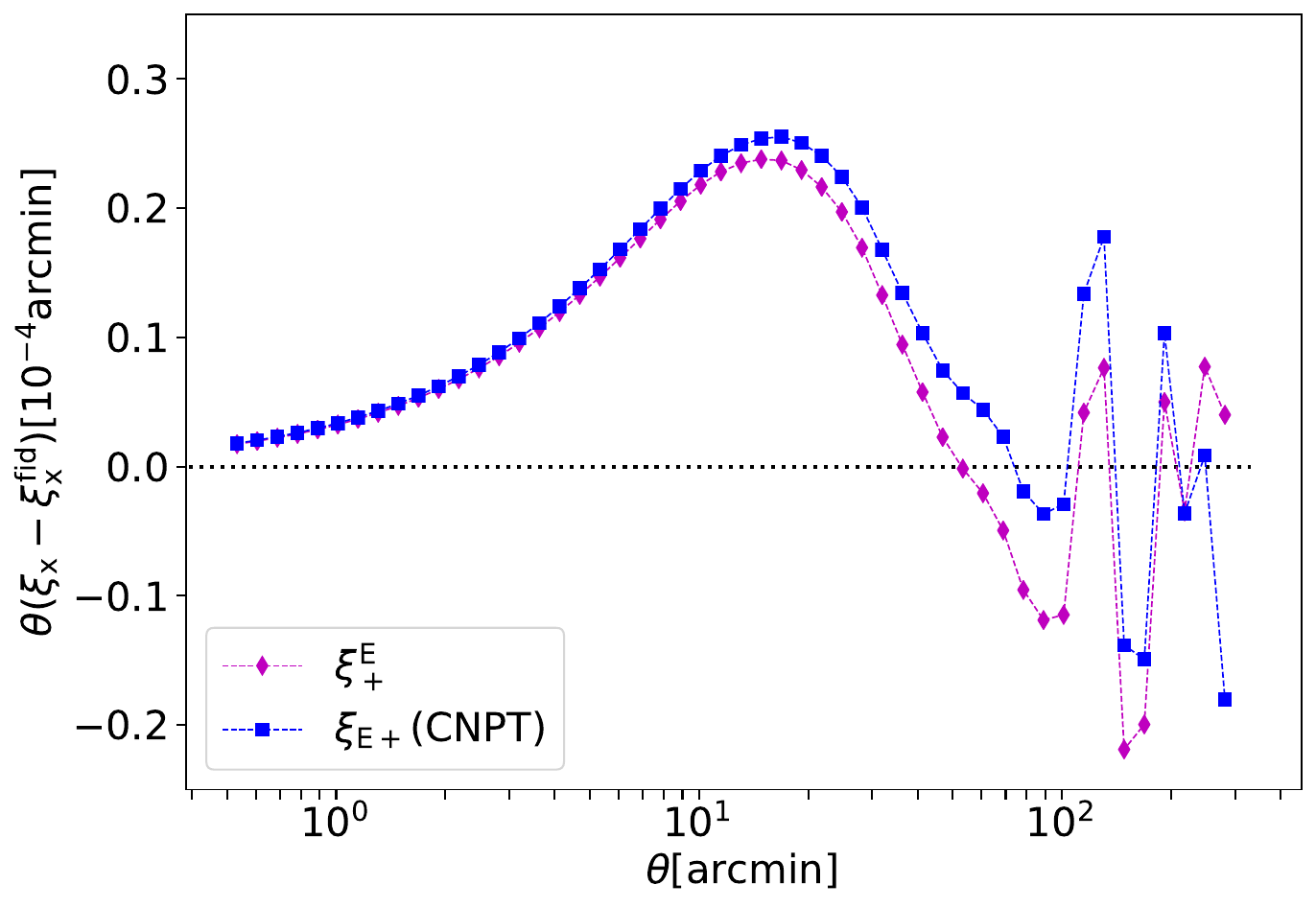}
  \includegraphics[width=\columnwidth]{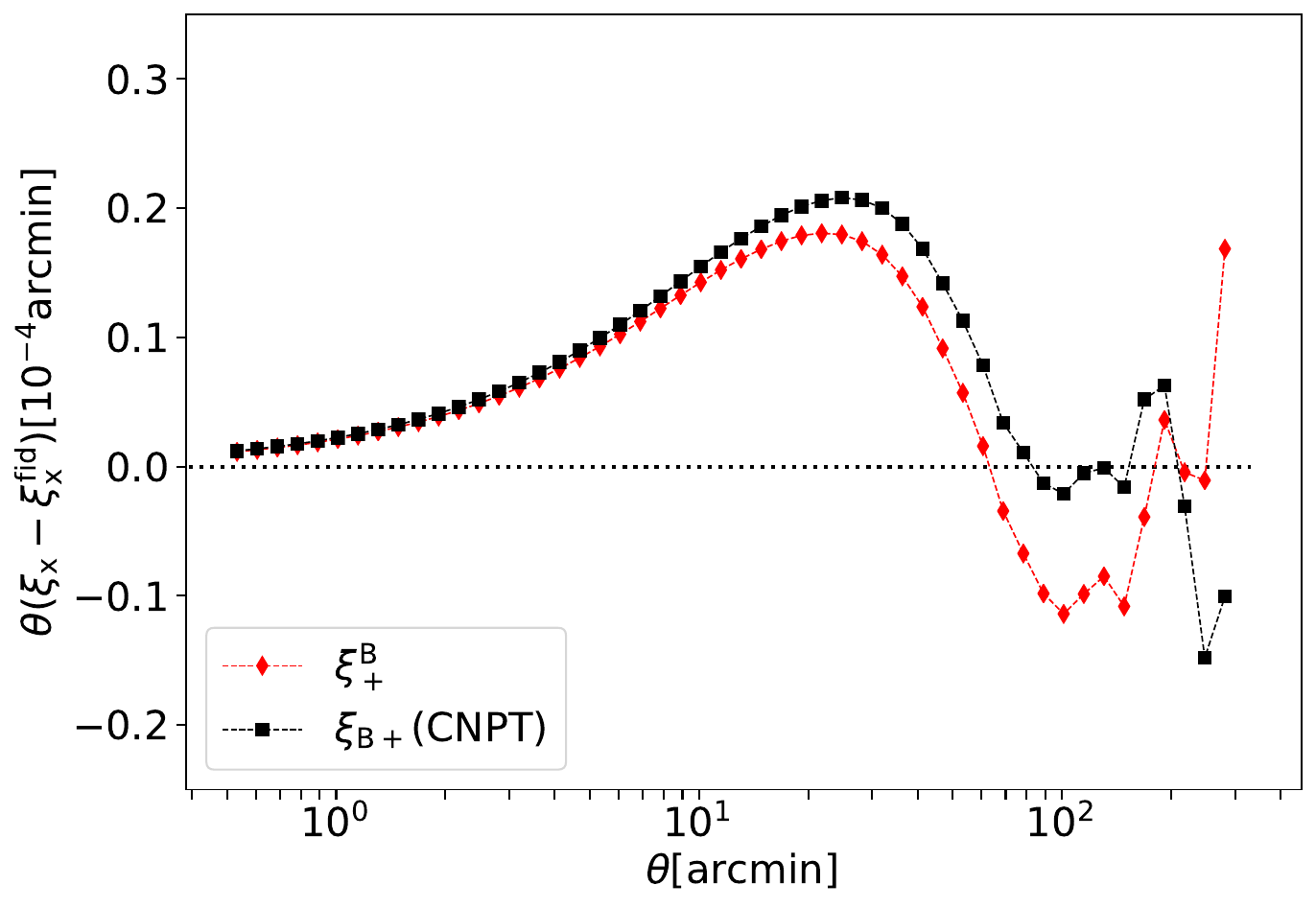}
  \caption{Comparison between the CNPT and pure-mode correlation
    functions on systematic-induced mock data, averaged over ten
    shape-noise-free lines-of-sight.  The point-spread function
    leakage as modeled by \cite{Asgari19a} is used here. The fiducial
    no-systematic signal is subtracted from the systematic-induced
    ones. All measurements are done for 50 logarithmic bins between
    0.5 and 300 arcminutes. }
  \flabel{slicspsf}
\end{figure*}

We chose the point-spread function leakage, as modeled in
\citet[][Sect. 5.1.1]{Asgari19a}, as a test case.  This systematics
introduces both, artificial E and B modes.
Figure\thinspace\ref{fig:slicspsf} illustrates the E- and B-mode
measurements in the left and right panels, respectively.  In all cases
the impact of the systematic is isolated via subtracting the fiducial
no-systematic signal shown in Fig.\thinspace\ref{fig:slicsfid}.  Again
the old and new measurements match at small $\theta$, while they
differ at larger scales.  The infinite upper bounds in
Eq.\,\eqref{eq:xiEBold} are more problematic here, since we do not
have a theoretical prediction for this systematic effect. Using the
pure-mode correlation functions allows us to isolate the scales where
systematic effects create B modes without the need for extrapolating
the measurements.

\subsection{\label{sc:Consistency}Consistency checks}
Having obtained explicit expressions for the pure-mode shear
correlation functions, we now apply two checks on their
consistency. First, we show explicitly that they are insensitive
to ambiguous modes. Second, we show that for a pure E-mode shear
field, the B-mode correlation functions vanish identically.

\subsubsection{Insensitivity of $\xi^{\rm E,B}_\pm$ to ambiguous modes}
As we mentioned before, some shear modes are neither E nor B modes,
and they should not affect the $\xi^{\rm E,B}_\pm$ functions. For
example, a constant shear field, with $\gamma(\vc\theta)=\gamma_0$
leads to a pair of correlation functions $\xi_+(\vt)=|\gamma_0|^2$,
$\xi_-(\vt)=0$. In this particular case, we find from
Eqs.\,(\ref{eq:xiplusE}, \ref{eq:xiplusB}) that
\be
\xi_+^{\rm E}(\vt)=\xi_+^{\rm B}(\vt)={1\over
  2}\,\eck{\xi_+(\vt)-S_+(\vt)}\;,
\elabel{xipEambig}
\ee
with all other terms vanishing. However, since 
\be
\int_\tmin^\tmax{\d\theta\;\theta\over \bar\vt^2} H_+(\vt,\theta)=1\;,
\ee
$S_+(\vt)=|\gamma_0|^2=\xi_+(\vt)$ and
$\xi_+^{\rm E}(\vt)=0=\xi_+^{\rm B}(\vt)$ in this case. Hence, this
ambiguous mode is filtered out. More generally, if we consider a
linear shear field, for which $\xi_+(\vt)=a+b(\vt/\bar\vt)^2$ and
$\xi_-(\vt)=0$, then again Eq.\,(\ref{eq:xipEambig}) holds, and since
\be
\int_\tmin^\tmax{\d\theta\;\theta\over \bar\vt^2} 
\eck{a+b\rund{\theta\over\bar\vt}^2}
H_+(\vt,\theta)=a+b\rund{\vt\over\bar\vt}^2
\;,
\ee
we again obtain $S_+(\vt)=\xi_+(\vt)$ and thus
$\xi_+^{\rm E}(\vt)=0=\xi_+^{\rm B}(\vt)$.

\subsubsection{$\xi^{\rm B}_\pm\equiv 0$ for pure E-mode shear}
As an important consistency check of the foregoing discussion, we now
want to show that the B-mode correlation functions $\xi^{\rm B}_\pm$
identically vanish if the shear field does not contain any B modes.
In this case, the two correlation functions $\xi_\pm$ are related
through
\begin{align}
\xi_+(\theta)&=\xi_-(\theta)+\int_\theta^\infty {\d\vp\over\vp}\;
\xi_-(\vp)\rund{4-{12\theta^2\over \vp^2}}\;, \nonumber \\
\xi_-(\theta)&=\xi_+(\theta)+\int_0^\theta{\d\vp\;\vp\over\theta^2}\;
\xi_+(\vp)\rund{4-{12 \vp^2\over \theta^2}} \;.
\elabel{xipmnoB}
\end{align}
Hence, in the absence of B modes, Eq.\,(\ref{eq:xiplusB}) reduces to
\be
2\xi^{\rm B}_+(\vt)=\int_\tmax^\infty {\d\theta\over\theta}\;
\xi_-(\theta)\rund{4-{12 \vt^2\over\theta^2}}-S_+(\vt)+S_-(\vt)\;.
\elabel{xiBplus1}
\ee
In order to show that this vanishes, we first consider the term $S_+$
and rewrite it with the help of Eq.\,(\ref{eq:xipmnoB}),
\begin{align}
S_+(\vt)&=\int_\tmin^\tmax{\d\theta\;\theta\over \bar\vt^2}\,
\xi_+(\theta)\,H_+(\vt,\theta)\nonumber\\
=& \int_\tmin^\tmax{\d\theta\;\theta\over \bar\vt^2}
\,H_+(\vt,\theta)
\bigg[ \xi_-(\theta)+\int_\theta^\tmax{\d\vp\over \vp}\xi_-(\vp)
\rund{4-{12\theta^2\over \vp^2}}\nonumber\\
&+\int_\tmax^\infty{\d\vp\over \vp}\xi_-(\vp)
\rund{4-{12\theta^2\over \vp^2}}\bigg] \nonumber \\
=&\int_\tmin^\tmax{\d\theta\;\theta\over \bar\vt^2}\,
H_+(\vt,\theta)\,\xi_-(\theta)
\elabel{Splus1}\\
&+\int_\tmin^\tmax{\d\vp\over\vp}\;\xi_-(\vp)
\int_\tmin^\vp {\d\theta\;\theta\over \bar\vt^2}\,H_+(\vt,\theta)\,
\rund{4-{12\theta^2\over \vp^2}}\nonumber\\
&+\int_\tmin^\tmax{\d\theta\;\theta\over \bar\vt^2}\,H_+(\vt,\theta)
\int_\tmax^\infty{\d\vp\over \vp}\xi_-(\vp)
\rund{4-{12\theta^2\over \vp^2}} \;,\nonumber
\end{align}
where the function $H_+(\vt,\theta)$ is given by
Eq.\,(\ref{eq:Hplus}), and in the second step we have changed the
order of integration, subject to the constraint
$\tmin\le\theta\le\vp\le\tmax$. Thus, we have rewritten $S_+$ solely
in terms of $\xi_-$, as are the other terms in
Eq.\,(\ref{eq:xiBplus1}). One finds that
\be
\int_\tmin^\tmax{\d\theta\;\theta\over \bar\vt^2}\,H_+(\vt,\theta)
\rund{4-{12\theta^2\over \vp^2}}
=4-{12\vt^2\over \vp^2} \;,
\ee
which shows that the final term in Eq.\,(\ref{eq:Splus1}) cancels the
first term on the r.h.s. of Eq.\,(\ref{eq:xiBplus1}). Hence, $\xi^{\rm
  B}_+$ does not have any contributions of $\xi_-$ from outside the
considered interval. The remaining terms are
\begin{align}
2\xi^{\rm
  B}_+(\vt)&=\int_\tmin^\tmax{\d\theta\over\theta}\;\xi_-(\theta)\;
\bigg[\rund{\theta\over\bar\vt}^2H_+(\vt,\theta) \nonumber \\
&+\int_\tmin^\theta{\d\vp\;\vp\over\bar\vt^2}
\,H_+(\vt,\vp)\rund{4-{12\vp^2\over \theta^2}}-H_-(\vt,\theta)\bigg]
\;, \elabel{xiBplus2}
\end{align}
where the function $H_-(\vt,\theta)$ is given by 
Eq.\,(\ref{eq:Hminus}). Carrying out the $\vp$ integral, one can show
that the bracket in Eq.\,(\ref{eq:xiBplus2}) vanishes identically, and
thus $\xi^{\rm  B}_+(\vt)\equiv 0$ in the absence of B modes.

Similarly, we find from Eqs.\,(\ref{eq:xiBminus}) and
(\ref{eq:xipmnoB}) in the case of vanishing B modes
\begin{align}
2\xi^{\rm B}_-(\vt)&=\int_\tmin^\tmax {\d\theta\;\theta\over
  \bar\vt^2}\,\xi_+(\theta) \,
\eck{K_-(\vt,\theta)-K_+(\vt,\theta)} \nonumber\\
&-\int_0^\tmin {\d\theta\;\theta\over
  \vt^2}\,\xi_+(\theta)\rund{4-{12\theta^2\over \vt^2}} 
\elabel{xiBmin1}\\
&+\int_\tmin^\tmax {\d\theta\;\theta\over
  \bar\vt^2}\,K_-(\vt,\theta)
\int_0^\theta{\d\vp\;\vp\over \theta^2}\,\xi_+(\vp)
\rund{4-{12\vp^2\over \theta^2}} \;. \nonumber
\end{align}
The last integral is then split into one from $0$ to $\tmin$ and
one from $\tmin$ to $\theta$. For the former, we note the result
that
\be
\int_\tmin^\tmax
{\d\theta\over\theta}\;K_-(\vt,\theta)\,\rund{4-{12\vp^2\over\theta^2}} 
={\bar\vt^2\over \vt^2}\rund{4-{12\vp^2\over\vt^2}}\;,
\ee
so that the corresponding  $\theta$ integral just cancels the
second term in Eq.\,(\ref{eq:xiBmin1}). Hence, $\xi^{\rm B}_-(\vt)$
contains no contribution from scales outside the angular interval
considered. For the $\theta$ integration of the second $\vp$ integral,
we change the order of integration, subject to
$\tmin\le\vp\le\theta\le\tmax$, to get
\begin{align}
2\xi^{\rm B}_-(\vt)&=\int_\tmin^\tmax {\d\theta\;\theta\over
  \bar\vt^2}\,\xi_+(\theta) \, \Bigg[
K_-(\vt,\theta)-K_+(\vt,\theta) \nonumber \\
&\qquad +\int_\theta^\tmax{\d\vp\over\vp}\;K_-(\vt,\vp)
\rund{4-{12\theta^2\over \vp^2}} \Bigg] \;.
\end{align}
One can show that the term in the bracket is identically zero, which
shows that  $\xi^{\rm B}_-(\vt)\equiv 0$ for the case that the shear
field has no B-mode contribution.

\subsection{\label{sc:ReltoP}Relation to the power spectrum}
We now consider the relation between the shear power spectra and the
pure-mode shear correlation functions. The $\xi_\pm(\vt)$ are related
to the E- and B-mode power spectra $P_{\rm E}(\ell)$ and
$P_{\rm B}(\ell)$ by
\begin{align}
\xi_+(\vt)&=\int_0^\infty{\d\ell\;\ell\over 2\pi}\;{\rm J}_0(\ell\vt)\,
\eck{P_{\rm E}(\ell)+P_{\rm B}(\ell)}\;, \nonumber \\
\xi_-(\vt)&=\int_0^\infty{\d\ell\;\ell\over 2\pi}\;{\rm J}_4(\ell\vt)\,
\eck{P_{\rm E}(\ell)-P_{\rm B}(\ell)}\; .
\elabel{xipmP}
\end{align}
Expressions (\ref{eq:xiplusE}, \ref{eq:xiplusB},
\ref{eq:xiEminus}, \ref{eq:xiBminus}) show that
$\xi_\pm^{\rm E/B}(\vt)$ are linear in the $\xi_\pm$ and hence can be
expressed in the form
\begin{align}
  \xi_\pm^{\rm E}(\vt)&=\int_0^\infty {\d\ell\;\ell\over
                2\pi}\,\eck{W_{\pm{\rm E}}^{\rm E}(\ell,\vt)\,P_{\rm
                E}(\ell)+W_{\pm{\rm B}}^{\rm E}(\ell,\vt)P_{\rm B}(\ell)}\;,
                        \nonumber \\
  \xi_\pm^{\rm B}(\vt)&=\int_0^\infty {\d\ell\;\ell\over
         2\pi}\,\eck{W_{\pm{\rm E}}^{\rm B}(\ell,\vt)\,P_{\rm
         E}(\ell)+W_{\pm{\rm B}}^{\rm B}(\ell,\vt)P_{\rm B}(\ell)} \;.
\end{align}
We start with $\xi_+^{\rm E}$, for which the coefficients read
\begin{align}
  W_{+{\rm E}}^{\rm E}&(\ell,\vt)={1\over 2}\Biggl[
  {\rm J}_0(\ell\vt)+{\rm J}_4(\ell\vt)                              
  +\int_\vt^\tmax{\d\theta\over\theta}\;{\rm J}_4(\ell\theta)\,
  \rund{4-{12 \vt^2\over \theta^2}} \nonumber \\
&-\int_\tmin^\tmax{\d\theta\;\theta\over \bar\vt^2}\,
{\rm J}_0(\ell\theta)\,H_+(\vt,\theta)
-\int_\tmin^\tmax{\d\theta\over\theta}\;{\rm J}_4(\ell\theta)\,
H_-(\vt,\theta)  \Biggr] \;, \nonumber \\
  W_{+{\rm B}}^{\rm E}&(\ell,\vt)={1\over 2}\Biggl[
  {\rm J}_0(\ell\vt)-{\rm J}_4(\ell\vt)                              
  -\int_\vt^\tmax{\d\theta\over\theta}\;{\rm J}_4(\ell\theta)\,
  \rund{4-{12 \vt^2\over \theta^2}} \nonumber \\
&-\int_\tmin^\tmax{\d\theta\;\theta\over \bar\vt^2}\,
{\rm J}_0(\ell\theta)\,H_+(\vt,\theta)
+\int_\tmin^\tmax{\d\theta\over\theta}\;{\rm J}_4(\ell\theta)\,
H_-(\vt,\theta)  \Biggr] \;. \nonumber
\end{align}
We expect that the latter coefficient vanishes, since the pure E-mode
correlation function should not depend on the B-mode power
spectrum. Indeed, it can be shown that
$W_{+{\rm B}}^{\rm E}(\ell,\vt)\equiv 0$. By adding the previous two
equations, we can simplify the expression for
$W_{+{\rm E}}^{\rm E}(\ell,\vt)$ to
\begin{align}
W_{+{\rm E}}^{\rm E}&(\ell,\vt)={\rm J}_0(\ell\vt)
-\int_\tmin^\tmax{\d\theta\;\theta\over \bar\vt^2}\,
{\rm J}_0(\ell\theta)\, H_+(\vt,\theta) \nonumber \\
&={\rm J}_0(\ell\vt)-{(1+B)\over 4 B^2 \ell\bar\vt}
\eck{3\rund{\vt\over \bar\vt}^2-(3-2B+3B^2)}
{\rm J}_1(\ell\tmax)\nonumber \\
&-{(1-B)\over 4 B^2 \ell\bar\vt}
\eck{3\rund{\vt\over \bar\vt}^2-(3+2B+3B^2)}
{\rm J}_1(\ell\tmin)\nonumber \\
&+{3 \over 4B^3 (\ell \bar\vt)^2}\,\eck{\rund{\vt\over
    \bar\vt}^2-(1+B^2)} \elabel{WpEE} \\
&\times\eck{ (1+B)^2\, {\rm J}_2(\ell\tmax)
-(1-B)^2\,{\rm J}_2(\ell\tmin) }  \;. \nonumber
\end{align}
We first note that the function $W_{+{\rm E}}^{\rm E}$ does not only
depend on the product $\ell\vt$, as was the case for the corresponding
filter for $\xi_+$. Since the pure-mode correlation functions depend
on the angular interval $\tmin\le\vt\le\tmax$, the filter
$W_{+{\rm E}}^{\rm E}$ has an explicit dependence on the interval
boundaries, expressed through $B$, $\bar\vt$ and the arguments of the
Bessel functions.  The additional terms in $W_{+{\rm E}}^{\rm E}$
filter out the ambiguous modes. In fact, since for small $\ell$,
$W_{+{\rm E}}^{\rm E}(\ell,\vt)\propto \ell^4$, low-$\ell$ modes in
the power spectrum are strongly suppressed.

The foregoing fact is an important observation. The filter that relates
$\xi_+$ to the power spectra is ${\rm J}_0(\ell\vt)$, which tends to
unity as $\ell\to 0$. Hence, $\xi_+$ is very sensitive to small-$\ell$
power (i.e., to large-scale modes). The fact that the filter
$W_{+{\rm E}}^{\rm E}$ has a leading $\ell^4$ dependence shows that
the sensitivity of $\xi_+$ to large-scale modes is due solely to the
ambiguous modes in $\xi_+$.

Turning to $\xi_+^{\rm B}$, it is straightforward to see that
$W_{+{\rm E}}^{\rm B}(\ell,\vt) = W_{+{\rm B}}^{\rm E}(\ell,\vt)=0$
and $W_{+{\rm B}}^{\rm B}(\ell,\vt)=W_{+{\rm E}}^{\rm
  E}(\ell,\vt)$. Thus, the pure B-mode correlation function is
independent of the E-mode power spectrum, and the relation between
$\xi_+^{\rm B}$ and $P_{\rm B}$ is the same as between $\xi_+^{\rm E}$
and $P_{\rm E}$.

The filter functions for $\xi_-^{\rm E}$ are
\begin{align}
  W&_{-{\rm E/B}}^{\rm E}(\ell,\vt)={1\over 2}\Biggl[
{\rm J}_0(\ell\vt)\pm {\rm J}_4(\ell\vt)
+ \int_\tmin^\vt {\d\theta\;\theta\over \vt^2}\,{\rm J}_0(\ell\theta)
\rund{4-{12\theta^2\over\vt^2}} \nonumber \\
&-\int_\tmin^\tmax{\d\theta\;\theta\over \bar\vt^2}\,
{\rm J}_0(\ell\theta)\,K_+(\vt,\theta)                          
\mp \int_\tmin^\tmax{\d\theta\;\theta\over \bar\vt^2}\;{\rm J}_4(\ell\theta)\,
K_-(\vt,\theta)  \Biggr] \;, \nonumber
\end{align}
where the upper (lower) signs apply for $W_{-{\rm E}}^{\rm E}$
($W_{-{\rm B}}^{\rm E}$). We find that
$W_{-{\rm B}}^{\rm E}(\ell,\vt)\equiv 0$, as expected, that is, the
B-mode power does not contribute to the pure E-mode correlation
function $\xi_-^{\rm E}$. Taking the sum of the two filter functions,
we find that
\begin{align}
  W_{-{\rm E}}^{\rm E}&(\ell,\vt)={\rm J}_0(\ell\vt)
  +\int_\tmin^\vt {\d\theta\;\theta\over \vt^2}\,{\rm J}_0(\ell\theta)
\rund{4-{12\theta^2\over\vt^2}} \nonumber \\
&-\int_\tmin^\tmax{\d\theta\;\theta\over \bar\vt^2}\,
        {\rm J}_0(\ell\theta)\,K_+(\vt,\theta) \nonumber \\
&={\rm J}_4(\ell\vt)
+{(1-B^2)\over 4 B^2 \ell\bar\vt} \eck{a_{-1}^{\rm min}{\rm J}_1(\ell\tmin)
+a_{-1}^{\rm max}{\rm J}_1(\ell\tmax)} \elabel{WmEE} \\
&+{3(1-B^2)^2\over 4 B^3 (\ell\bar\vt)^2}
\eck{a_{-2}^{\rm min}{\rm J}_2(\ell\tmin)
+a_{-2}^{\rm max}{\rm J}_2(\ell\tmax)} \;, \nonumber
\end{align}
where the coefficients are
\begin{align}
  a_{-1}^{\rm min}&=(1+B)\eck{3(1-B^2)^2\rund{\vt\over\bar\vt}^{-4}
  -(3-2B+3B^2)\rund{\vt\over\bar\vt}^{-2}} \;, \nonumber\\
  a_{-1}^{\rm max}&=(1-B)\eck{3(1-B^2)^2\rund{\vt\over\bar\vt}^{-4}
  -(3+2B+3B^2)\rund{\vt\over\bar\vt}^{-2}} \;,\nonumber\\
  a_{-2}^{\rm min}&=(1+B)^2(1-4B+B^2)\rund{\vt\over\bar\vt}^{-4}
  -(1-B^2)\rund{\vt\over\bar\vt}^{-2} \;, \\
  a_{-2}^{\rm max}&=(1+B^2)\rund{\vt\over\bar\vt}^{-2}
  -(1-B)^2(1+4B+B^2)\rund{\vt\over\bar\vt}^{-4} \;.\nonumber
\end{align}
Finally, we find $W_{-{\rm E}}^{\rm B}(\ell,\vt)\equiv 0$, again as
expected since the correlation function $\xi_-^{\rm B}(\vt)$ should
not depend on the E-mode power spectrum, and
$W_{-{\rm B}}^{\rm B}(\ell,\vt)=W_{-{\rm E}}^{\rm E}(\ell,\vt)$. Thus,
of the eight filter functions $W_{\pm{\rm E/B}}^{\rm E/B}$, four are
identically zero, and the remaining four are pairwise identical, so
that only the two given in Eqs.\,(\ref{eq:WpEE}) and (\ref{eq:WmEE})
need to be evaluated.

We note that as $\tmin\to 0$, $\tmax\to\infty$,
$W_{+{\rm E}}^{\rm E}(\ell,\vt)\to {\rm J}_0(\ell\vt)$ and
$W_{-{\rm E}}^{\rm E}(\ell,\vt)\to {\rm J}_4(\ell\vt)$, due to the
behavior of the Bessel functions for small and large arguments. Hence,
in this case the relation between the pure-mode shear correlation
functions and the power spectra reduces to that of the CNPT
correlation functions.

Finally, from the decomposition (\ref{eq:decomposition}) of the
correlation functions and the results of this subsection, we find
the relation between the ambiguous modes and the power spectra,
\begin{align}
\xi_+^{\rm amb}(\vt)&=\int{\d\ell\over 2\pi}\,
\eck{{\rm J}_0(\ell\vt)-W_{+\rm E}^{\rm E}(\ell,\vt)}\,
\eck{P_{\rm E}(\ell)+P_{\rm B}(\ell)}\;,\nonumber \\
\xi_-^{\rm amb}(\vt)&=\int{\d\ell\over 2\pi}\,
\eck{{\rm J}_4(\ell\vt)-W_{-\rm E}^{\rm E}(\ell,\vt)}\,
\eck{P_{\rm E}(\ell)-P_{\rm B}(\ell)}\;.
\end{align}
Given that both of the $\xi_\pm^{\rm amb}(\vt)$ are characterized by
only two coefficients, it is obvious that one can find many
combinations of E- and B-mode power spectra for which these coefficients
are the same. Therefore, these ambiguous mode correlation functions
can result from different combinations of E and B modes. We give some
specific examples for this in Appendix \ref{sc:AppA3}.

\section{\llabel{kids} KiDS-1000 measurements}

\begin{figure*}
  \includegraphics[width=2\columnwidth]{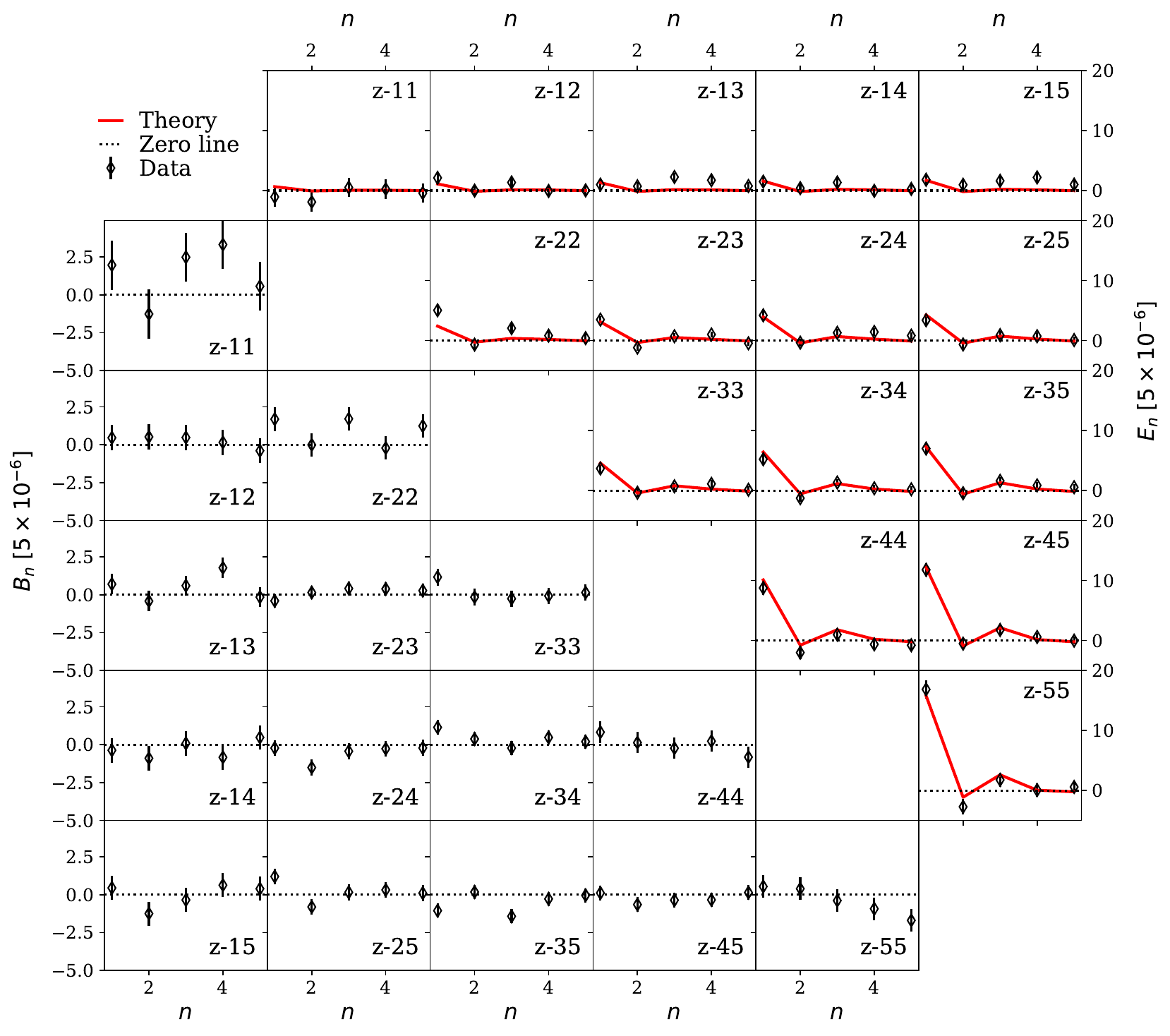}
  \caption{Dimensionless logarithmic COSEBI (see Appendix
    \ref{sc:newCOSEBI}) measurements from KiDS-1000 data. The
    E and B modes are shown in the top and bottom triangles, respectively. Each panel
    depicts results for a pair of redshift bins, $z$-$ij$. The solid
    red curves correspond to the best fitting model to the SEK COSEBIs
    as analyzed in \citet[compare with their Fig. 3]{Asgari21}. The
    B modes are consistent with zero ($p$-${\rm value}=0.36$) and the
    best-fit model describes the data very well
    ($p$-${\rm value}=0.2$). We note that the COSEBI modes are discrete
    and the points are connected to one another for visual aid. }
  \flabel{kidscosebis}
\end{figure*}

\begin{figure*}
  \includegraphics[width=2\columnwidth]{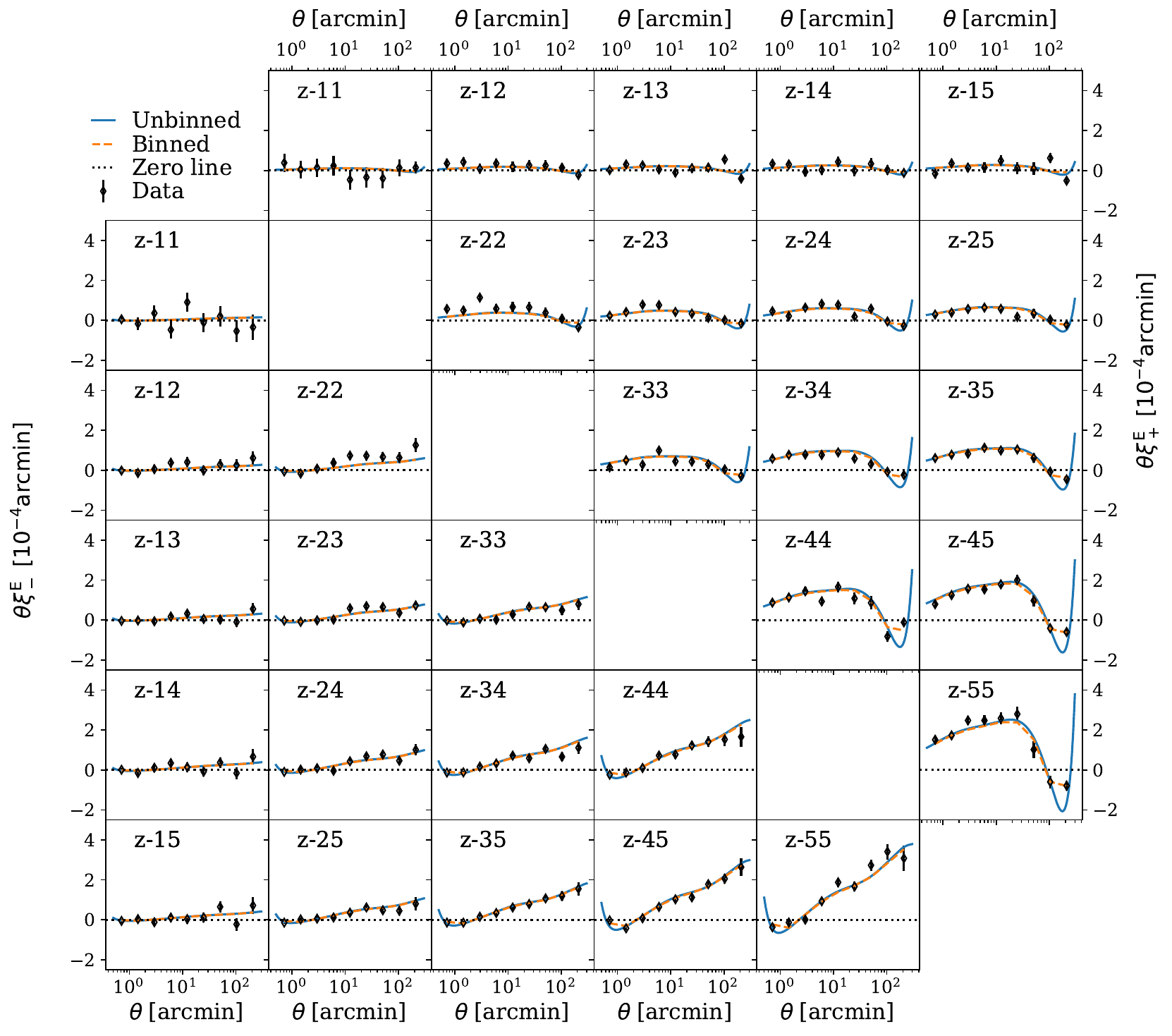}
  \caption{KiDS-1000 pure E-mode correlation functions. The top and
    bottom panels show $\xi_+^{\rm E}$ and $\xi_-^{\rm E}$,
    respectively. The theory curve is shown for both unbinned (solid blue) and binned (dashed orange) cases. The data points should be
    compared with the binned curve. The model is calculated assuming
    the best fitting standard cosmology to SEK COSEBIs
    \citep{Asgari21}.  Although the model is not fitted to this data
    vector, we find that it agrees with the data very well
    ($p$-${\rm value}=0.09$ for $\xi_{+}^{\rm E}$ and 0.28 for
    $\xi_{-}^{\rm E}$).}
  \flabel{kidsxiE}
\end{figure*}

\begin{figure*}
  \includegraphics[width=2\columnwidth]{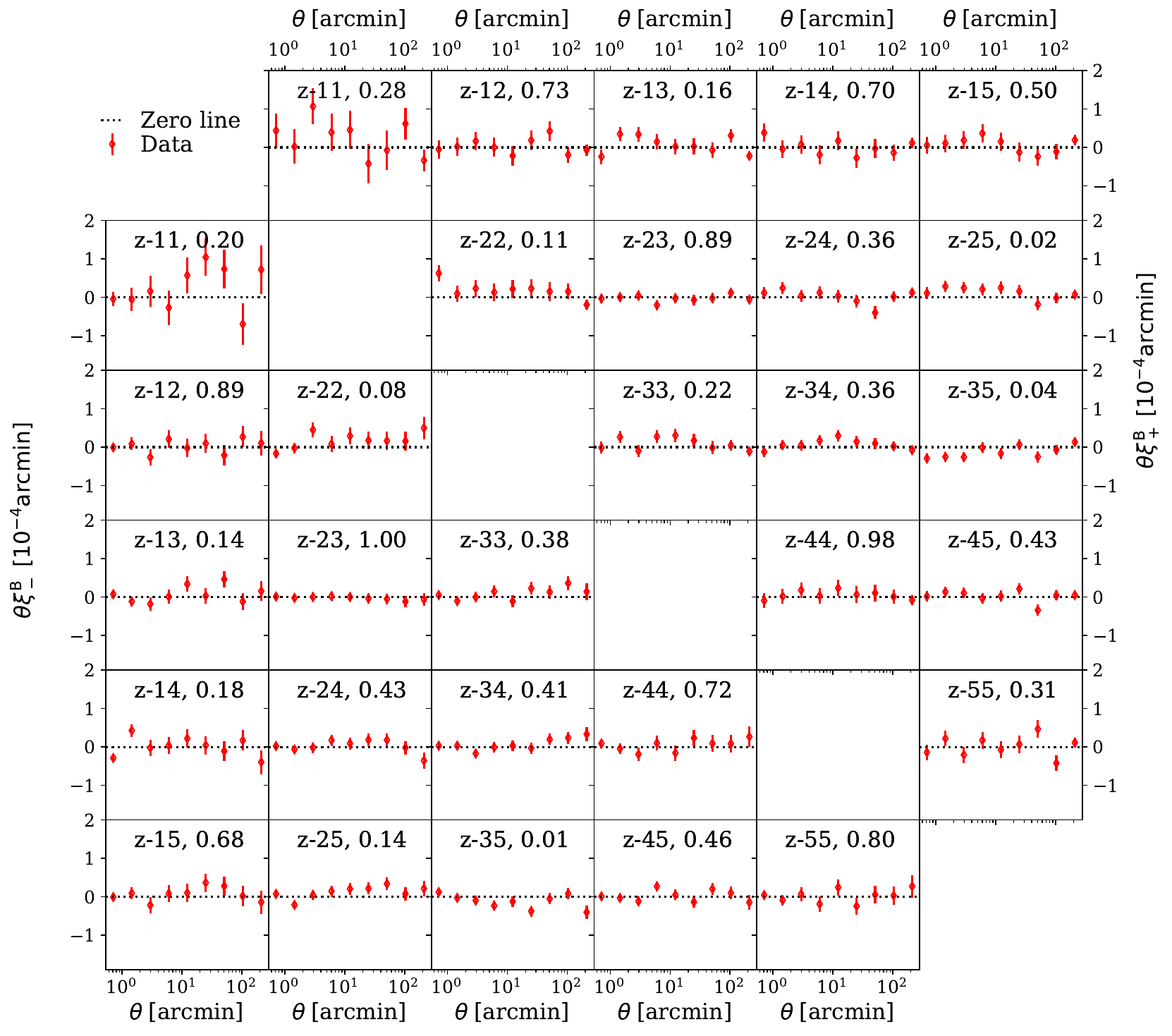}
  \caption{KiDS-1000 pure B-mode correlation functions.
    $\xi_+^{\rm B}$ is displayed in the top panels, while
    $\xi_-^{\rm B}$ is shown in the bottom ones. Each panel represents
    measurements for a pair of redshift bins, $z-ij$, and its associated
    $p$-value.  We find that the B modes are consistent with zero when
    we consider the full data vectors ($p$-${\rm value}=0.11$ for
    $\xi_{+}^{\rm B}$ and 0.20 for $\xi_{-}^{\rm B}$).}
  \flabel{kidsxiB}
\end{figure*}

\subsection{Data description}
The Kilo Degree Survey is designed with weak gravitational lensing
applications in mind. Its data, therefore, benefit from high-quality
images in the $r$-band (mean seeing of 0.7 arcseconds), which is used
for the shape measurements \citep{Giblin21}.  In addition, all
galaxies have matched depth images in optical, $ugri$, and
near-infrared photometric bands, $ZYJHK_{\rm s}$.  The five
near-infrared bands are observed by the VISTA
Kilo-degree INfrared Galaxy (VIKING) survey \citep[][]{Edge13}.  These nine bands are
used to estimate photometric redshifts for all galaxies that
contribute to the cosmic shear signal.  The fourth KiDS data release
includes 1006 square degrees of images \citep{Kuijken19}.  The data are
divided into five tomographic bins before 2PCFs are measured for the 15 distinct combinations of redshift
bins\footnote{https://github.com/KiDS-WL/Cat\_to\_Obs\_K1000\_P1}. The
redshift distribution for each tomographic bin is calibrated using
KiDS+VIKING-like observations of fields containing spectroscopic
samples \citep{Hildebrandt21}.

The theoretical predictions were calculated with the KiDS Cosmology
Analysis Pipeline\footnote{https://github.com/KiDS-WL/kcap} ({\sc
  kcap}), which is built from the modular cosmology pipeline {\sc
  CosmoSIS}\footnote{https://bitbucket.org/joezuntz/cosmosis/wiki/Home}
\citep{Zuntz15}. The primordial power spectrum was estimated using the
{\sc camb} Boltzmann code \citep{camb}. Its nonlinear evolution was
calculated via the augmented halo model approach of \cite{Mead15},
which also accounts for the impact of baryon feedback from active
galactic nuclei. We modeled the intrinsic alignments of galaxies with
the nonlinear linear alignment (NLA) model of \citet[see also Hirata
\& Seljak \citeyear{Hirata_Seljak04}]{Bridle-King07} and used a
modified Limber approximation \citep{LoVerde-Afshordi08} to project
the three-dimensional power spectra into two dimensions,
$P_{\rm E}(\ell)$. This was then used to make predictions for the pure
mode correlation functions and the new dimensionless COSEBIs.

\subsection{COSEBIs and pure-mode correlations for KiDS-1000}
We calculated the new dimensionless logarithmic COSEBIs (see Appendix
\ref{sc:newCOSEBI}) by integrating over the measured
$\xi_\pm$.\footnote{We refer the reader to \cite{Asgari17} for details
  on this conversion from $\xi_\pm$ to COSEBIs.} The pure-mode
correlation functions were determined by integrating over the
$\xi_\pm$, according to the relations given in
Sect.\,\ref{sc:PMfromxi}. As a consistency check, we also calculated
$\xi_\pm^{\rm E/B}$ using Eqs.\,(\ref{eq:xiEBplus}) and
(\ref{eq:xiEBminus}), using the first 20 COSEBIs modes. We found that
the sums in Eqs.\,(\ref{eq:xiEBplus}) and (\ref{eq:xiEBminus})
converge to the previous result after about the first five COSEBI modes.

Figures\,\ref{fig:kidscosebis}, \ref{fig:kidsxiE}, and
\ref{fig:kidsxiB} display the measured dimensionless COSEBIs,
$\xi_\pm^{\rm E}$ and $\xi_\pm^{\rm B}$ for the angular separation
range of 0.5 to 300 arcminutes. In these figures, the error bars are
drawn from the diagonal of their respective covariance matrix.  Each
panel belongs to a pair of redshift bins. The theoretical curves were
calculated using the best fitting flat $\Lambda$CDM cosmology to the
KiDS-1000 cosmic shear data \citep[SEK COSEBIs;][]{Asgari21} whose
parameter values are given in Table\,\ref{tab:cosmoparams}. Although
not listed here we also fix the mean redshift displacement parameters
to their best fitting values as estimated in \cite{Asgari21}.  In all
cases, the theory values are connected to each other for ease of
comparison, although they are all discrete with the exception of the
unbinned theory curves (blue) in Fig\,\ref{fig:kidsxiE}. For COSEBIs
this is true by definition, while for $\xi_\pm^{\rm E/B}$ the binning
of the data requires the theoretical predictions to also be binned
(orange dashed curves).

\subsection{Covariances and Fisher analysis}
We first derived the covariance matrix for the new COSEBIs using the
methodology detailed in \cite{joachimi21} and Appendix A of
\cite{Asgari20}. The corresponding correlation matrix is shown in the
left panel of Fig.\,\ref{fig:corr} and compared to the correlation
matrix for the SEK COSEBIs shown on the right. As can be seen, the
dimensionless COSEBIs are considerably less correlated, making them
more mutually independent.

\begin{figure*}
  \includegraphics[width=2\columnwidth]{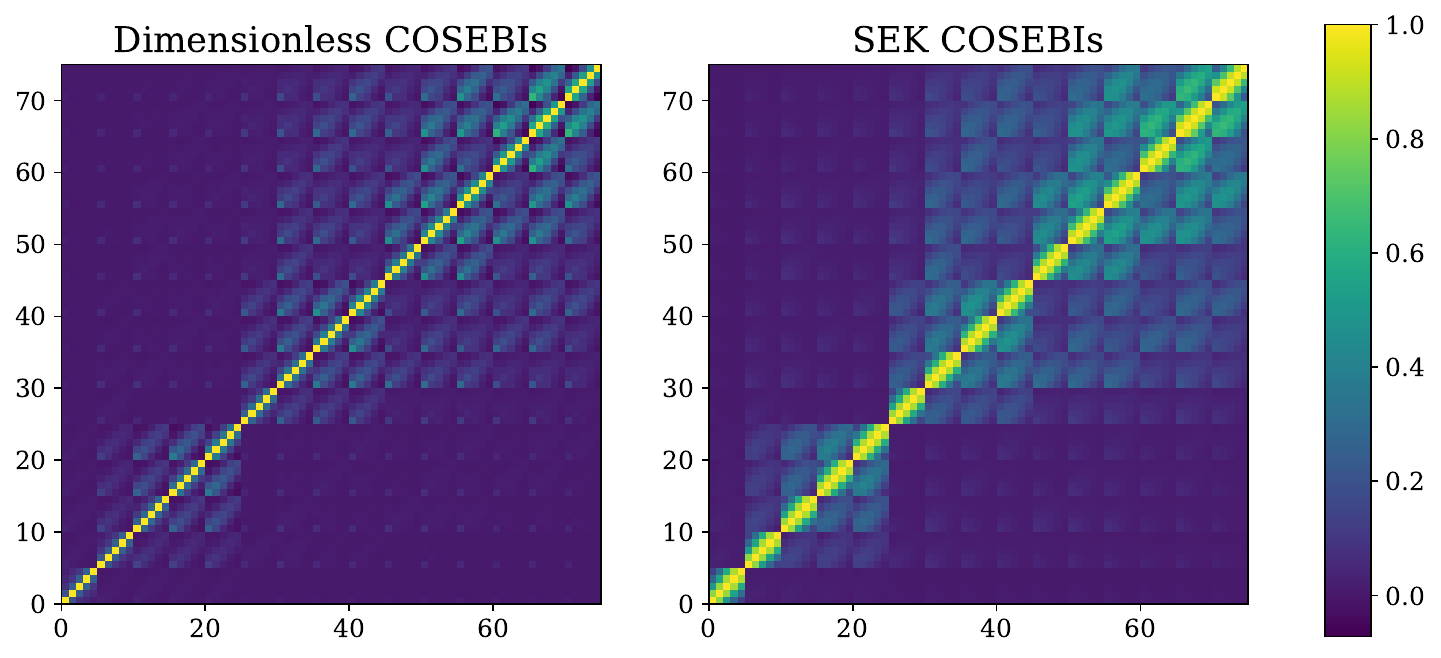}
  \caption{Correlation matrices for new (left) and old (right)
    logarithmic COSEBIs.  Here we illustrate the correlation matrices
    for the first five COSEBI modes.  Each five-by-five block shows the
    values for one pair of redshift bins, starting with the lowest
    bins at the bottom-left corner. }
  \flabel{corr}
\end{figure*}

\begin{figure*}
  \includegraphics[width=2\columnwidth]{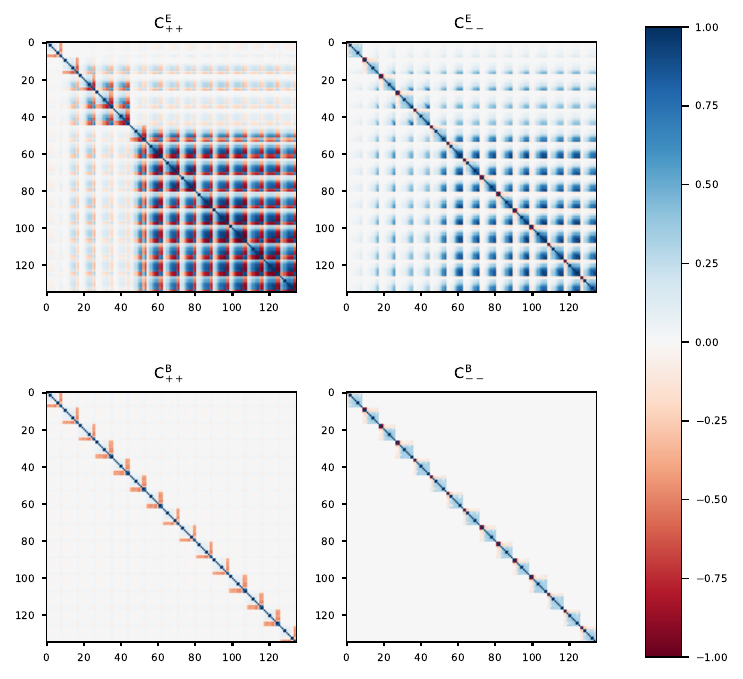}
  \caption{Correlation coefficients for pure-mode correlation
    functions. They are shown for the autocorrelations of
    $\xi^{\rm E}_+$ (top left), $\xi^{\rm E}_-$ (top right),
    $\xi^{\rm B}_+$ (bottom left), and $\xi^{\rm B}_-$ (bottom
    right). The covariance matrices are calculated for nine $\theta$ bins
    and five redshift bins, resulting in 15 distinct pairs of
    redshifts. The top-left corner of each panel shows the correlation
    coefficients for the lowest-redshift bins.}
  \flabel{corrxiEB}
\end{figure*}

We then estimated the covariance matrices for the pure-mode correlation
functions, making use of the linear relation between them and the
COSEBIs given by Eqs.\,\eqref{eq:xiEBplus} and
\eqref{eq:xiEBminus}. The correlation coefficients are shown in
Fig.\,\ref{fig:corrxiEB} for $\xi_\pm^{\rm E/B}$.

Although the theoretical curves are not fitted to the data in
Figs.\,\ref{fig:kidscosebis} and \ref{fig:kidsxiE}, we see that they
describe the data very well.\footnote{In principle, as mentioned
  before, the dimensionless COSEBIs and the pure-mode correlation
  functions should yield exactly the same result as using the SEK
  COSEBIs, as all these quantities contain the same information. In
  practice, however, the results will slightly differ, due to the use
  of a finite number of COSEBI modes and a finite number of $\vt$ bins
  for the correlation functions.} We estimated the goodness-of-fit
using the probability of exceeding the measured $\chi^2$ (i.e., the
$p$-value). Following \cite{joachimi21} we assume that the effective
number of free parameters is 4.5 and set the degrees of freedom to the
number of data points, minus 4.5. We then find that all $p$-values are
above 0.09 ($p$-values for each data vector are reported in the
caption of their figure). This is to be expected as the fit is done to
the SEK COSEBIs ($p$-value = 0.16), which separate E and B modes on the
same angular range.  Figure\,\ref{fig:kidsxiB} and the bottom panels
of Fig.\,\ref{fig:kidscosebis}, depict the B-mode signals. We find
that the B modes are consistent with zero in all cases and all
$p$-values are above 0.1. We also report the $p$-values for individual
pairs of redshift bins in Fig.\,\ref{fig:kidsxiB}, which can be
compared with the results of \cite{Giblin21} who used SEK COSEBIs to
determine the significance of B modes in KiDS-1000 data. We note that, as
demonstrated in \cite{Asgari19a}, the significance of the B modes has a
nontrivial dependence on the way the data are binned and, equivalently, on
the number of COSEBI modes that are used\footnote{The number of COSEBI
  modes and theta bins do not have a one-to-one relation. However, the
  higher COSEBI modes are more sensitive to smaller scale variations
  across the full range of the correlation functions. These variations
  are lost when data are binned coarsely.}, as well as on the types of
systematic effects that exist in the data. While certain systematics
produce E and B modes on similar angular separations (see for example
the impact of point-spread-function leakage in
Fig.\,\ref{fig:slicspsf}), others such as 
a CCD-chip bias that produces a repeating pattern in the images
\citep[see for example][regular pattern Figs.\,9 and 10]{Asgari19a},
show a different scale dependence for E and B modes. Therefore,
similar to COSEBIs here we recommend to use multiple binning schemes
to test the significance of B modes. In fact, we found similar trends
to \cite{Giblin21} depending on the number of $\theta$ bins. When we
divide the $[0\arcminf5,300']$ range into 20 $\theta$ bins we found that bin
55 has the smallest $p$-${\rm value}=0.04$, whereas dividing the same
range into five bins resulted in smaller $p$-values for redshift bin
combinations 22 and 35. Nevertheless, all $p$-values are above the
0.01 threshold and thus we conclude that the B modes are
insignificant. We also found that by increasing the number of $\theta$
bins, the $p$-values resulting from $\xi_+^{\rm B}$ and
$\xi_-^{\rm B}$ become very similar, confirming that these two
functions contain the same information.

We compare the information content of the pure mode correlation
function, $\xi_+^{\rm E}$ with the SEK COSEBIs, in
Fig.\,\ref{fig:fisher}.  We use a Fisher formalism and assume the
fiducial values in Table\,\ref{tab:cosmoparams} for model parameters.
As was shown in \cite{Asgari21}, we expect to have meaningful
constraints only for the structure growth parameter $S_8$ and the
amplitude of the intrinsic alignments $A_{\rm IA}$.  Therefore, we fixed
all other parameters and only show the 1$\sigma$ and 2$\sigma$ contours for
$S_8$ and $A_{\rm IA}$.  We see that the information content of
$\xi_+^{\rm E}$ and COSEBIs is identical, and conclude that there is
no extra cosmological information to be gained from the pure-mode
correlation functions. This is also true when we compare the
dimensionless and SEK COSEBIs Fisher matrices.  This is to be
expected, as both methods make use of the E-mode information that is
available in the given angular interval.  With the Fisher analysis we
can also estimate the expected errors on the measured parameters. We
find that the error on $S_8$ is 0.014 and on $A_{\rm IA}$ is 0.274,
both are slightly smaller than the full likelihood analysis of
\cite{Asgari21}, as expected.

\begin{figure}
  \includegraphics[width=\columnwidth]{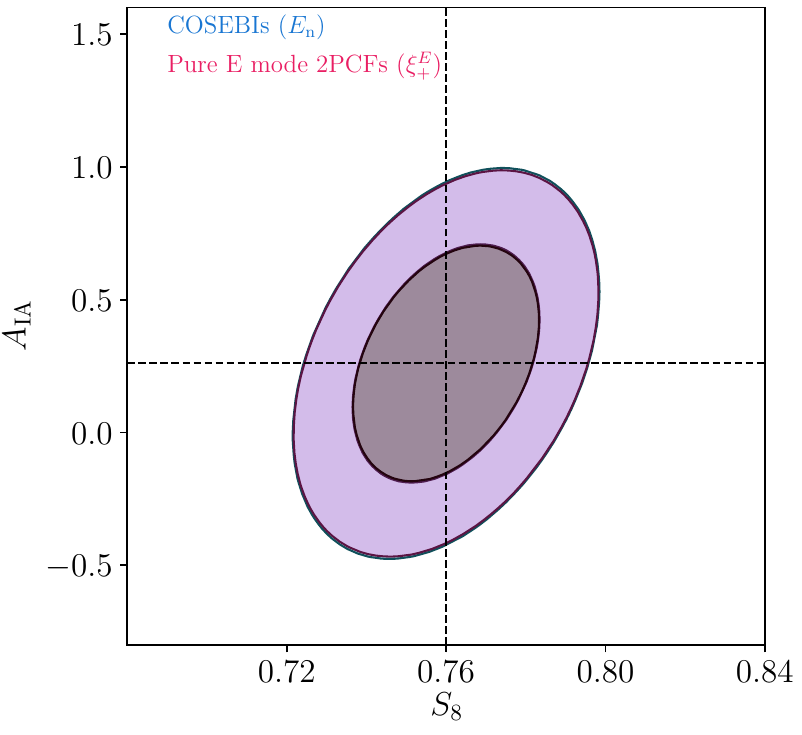}
  \caption{Fisher matrix forecast for KiDS-1000. The SEK COSEBIs
    (blue) are compared with $\xi^{\rm E}_+$ (pink), showing that they
    contain the same level of information about the model parameters,
    $S_8$ and $A_{\rm IA}$. All other parameters, listed in
    Table\,\ref{tab:cosmoparams}, are fixed to their fiducial
    values. The fact that one sees only one ellipse is because both
    methods give the same Fisher ellipses, which thus lie on top of each
    other, as expected. This figure is made with {\sc ChainConsumer}
    \citep{Hinton16}. }
  \flabel{fisher}
\end{figure}

\section{\llabel{Summary} Summary and discussion}
In this paper we have derived pure-mode shear correlation functions that
can be obtained from the measured shear correlations, $\xi_\pm(\vt)$, on
a finite interval, $0<\tmin\le\vt\le\tmax<\infty$. This was achieved by
redefining the orthonormality relation of COSEBIs, which allowed us
to construct two complete sets of orthonormal weight functions,
$T_{\pm \mu}(\vt)$, on this finite interval; explicit expressions for
these new weight functions are given in Appendix \ref{sc:newCOSEBI}.
Two of these weight functions correspond to ambiguous modes, and with
the remaining ones, the mode-separating COSEBIs were
defined. Owing to the completeness of these function sets, we were able to
decompose the shear correlation functions into their E- and B-mode
correlations, $\xi_\pm^{\rm E/B}(\vt)$, and their contribution by
ambiguous modes (see Eq.\,\ref{eq:decomposition}). These different
components can be straightforwardly determined from the
$\xi_\pm( \vt)$ measured on a finite interval, in contrast to the CNPT
correlation functions (see Sect.\,\ref{sc:oldxis}), which require extrapolation or the modeling of $\xi_\pm$ for separations smaller or
larger than where measurements of $\xi_\pm$ are available. Hence,
there is no longer any reason to use these CNPT correlation
functions. Only in the limit of $\tmin\to 0$ and $\tmax\to\infty$ do
they agree with mode-separating ones.

These new correlation functions allow the study of E- and B-mode
second-order shear as a function of angular scale. Hence, they should
serve as a diagnostic for the angular dependence of potential B modes
in a survey. To illustrate this, we applied the pure-mode
correlation functions to simulation data, without and with systematics
added, and compared the results with our earlier analysis
\citep{Asgari19a}.

We applied the newly constructed dimensionless COSEBIs to the
KiDS-1000 tomographic cosmic shear data set, for which we also
computed the pure-mode shear correlation functions. Calculating the
covariance of the COSEBIs and the binned $\xi_\pm^{\rm E/B}$, we
have shown that their measured values are fully consistent with the
best fitting model parameters obtained in \cite{Asgari21}, exhibiting
only very small differences in the $p$-values.  Using the Fisher
analysis, we also showed that the results on the two parameters best
constrained by the cosmic shear data ($S_8$ and $A_{\rm IA}$) are
indistinguishable between the COSEBIs and the pure-mode correlation
functions -- as was to be expected. The discrete nature of the COSEBIs
makes them the more convenient quantities for a cosmological analysis.

In Appendix\,\ref{sc:Ambig} we provide a few illustrative examples
of ambiguous modes in the shear correlation function, that is, modes
that cannot be uniquely attributed either to E or
B modes. Incorporation of such modes into a cosmological analysis
carries the risk that they are affected by a contribution coming
from B modes, and hence the analysis may be biased. We therefore
caution against the use of ambiguous modes when deriving constraints on
model parameters; instead, employing COSEBIs for that purpose avoids
this potential trap.  We note that the sensitivity of $\xi_+(\vt)$ to
low-$\ell$ power, due to the filter ${\rm J}_0(\ell\vt)$ relating
them, is solely due to ambiguous modes; the corresponding filter
function for the pure-mode correlation has an $\ell^4$ dependence for
$\ell\to 0$.

As was shown in \cite{Asgari12}, if one assumes that the ambiguous
modes are pure E modes, then they contain additional cosmological
information -- this corresponds to the case termed ``full COSEBIs'' in
\cite{Asgari12}. The relative amount of information in these ambiguous
modes depends on the angular range $\tmin$ to $\tmax$, and presumably
on the number of cosmological parameters. However, as was made clear
above, from the measurement of the correlation functions on this
finite interval, one cannot tell whether these ambiguous modes are
pure E modes or whether B modes are mixed in. We therefore strongly
advise against the use of ambiguous modes for cosmological parameter
inference.

The same statement holds for the correlation functions $\xi_\pm$;
to use them for cosmological parameter estimates, one needs to
(implicitly) assume that they are pure E-mode functions, which cannot
be verified from a measurement on a finite angular separation
interval. Thus, such estimates may contain an unknown level of
systematics due to B modes that remain undetected by the COSEBIs but
are hidden in the ambiguous modes.

Finally, we show in Appendix \ref{sc:subCOSEBI} that the COSEBIs
defined on a subinterval of the original one can be obtained as
linear combinations of the original COSEBIs. This was to be expected
since these original COSEBIs contain the full E and B mode-separable
information about second-order shear statistics. We thus conclude that
it suffices to consider the COSEBIs on the full angular range where
the $\xi_\pm$ are measured without needing to consider
subintervals. The lack of localized information in the individual
COSEBIs is remedied by the use of the pure-mode shear correlation
functions derived here.

\begin{acknowledgement}
  We acknowledge the constructive comments by the anonymous referee
  which led to an improvement of the presentation.  This work was
  supported by the Deutsche Forschungsgemeinschaft with the grant
  SCHN342-13 and the Heisenberg grant Hi 1495/5-1, the European
  Research Council under grants number 647112 and 770935, by an STFC
  Ernest Rutherford Fellowship (project reference ST/S004858/1), by
  the Max Planck Society and the Alexander von Humboldt Foundation in
  the framework of the Max Planck-Humboldt Research Award endowed by
  the Federal Ministry of Education and Research ERC with the
  Consolidator Grant No. 770935, by the Vici grant 639.043.512,
  financed by the Netherlands Organisation for Scientific Research
  (NWO), by the Royal Society and Imperial College, by the
  CMS-CSST-2021-A01, NSFC of China under grant 11973070, the Shanghai
  Committee of Science and Technology grant No.19ZR1466600 and Key
  Research Program of Frontier Sciences, CAS, Grant No. ZDBS-LY-7013,
  and the Leverhulme Trust.  Based on observations made with ESO
  Telescopes at the La Silla Paranal Observatory under programme IDs
  177.A-3016, 177.A-3017, 177.A-3018 and 179.A-2004, and on data
  products produced by the KiDS consortium. The KiDS production team
  acknowledges support from: Deutsche Forschungsgemeinschaft, ERC,
  NOVA and NWO-M grants; Target;
  the University of Padova, and the University Federico II (Naples).\\  
  {\it Author contributions}: All authors contributed to the
  development and writing of this paper. The authorship list is given
  in two groups: the lead authors (PS,MA,YNJ) followed by an
  alphabetical group that covers those who have either made a
  significant contribution to the data products, or to the scientific
  analysis.
\end{acknowledgement}

\bibliographystyle{aa}
\bibliography{WL}

\clearpage

\begin{appendix}
\section{\label{sc:Ambig}Shear fields from ambiguous modes}
In this appendix, we consider ambiguous modes of the shear field
in more detail. This will be done in different ways. First, we
give several examples of shear fields that cannot uniquely be
assigned to either E-mode or B-mode shear. We then show that a
statistical ensemble of such shear fields give rise to the ambiguous
modes in the shear 2PCFs. Finally, we show
that ambiguous modes in the shear correlation functions can be caused
by various combinations of E- and B-mode power spectra.

\subsection{\label{sc:AppA1}Ambiguous shear fields}
Following \cite{Crittenden02} and \cite{SvWM02}, we formally describe a
general shear field by a superposition of E and B modes, by defining
the complex deflection potential $\psi(\vc\theta)=\psi^{\rm
E}(\vc\theta)+{\rm i} \psi^{\rm B}(\vc\theta)$, where $\psi^{\rm
E/B}$ are real functions. The corresponding convergence is then
obtained from
the Poisson equation, $\kappa(\vc\theta)=\kappa^{\rm
  E}(\vc\theta)+{\rm i}\kappa^{\rm B}(\vc\theta)=(1/2)
\nabla^2\psi(\vc\theta)$. The shear field is given by
\be
\gamma=\gamma_1+{\rm i}\gamma_2=
\eck{{\psi_{,11}^{\rm E}-\psi^{\rm E}_{,22}\over 2}
  -\psi^{\rm B}_{,12}}
+{\rm i}\eck{\psi^{\rm E}_{,12} +{\psi_{,11}^{\rm B}
    -\psi^{\rm B}_{,22}\over 2}}\;,
\ee
where subscripts following a comma denote partial derivatives with
respect to $\theta_i$. We consider the following combinations of
second derivatives of the shear,
\begin{align}
  C_{\rm c}&:=\gamma_{2,11}-\gamma_{2,22}-2\gamma_{1,12}
  ={1\over 2}(\psi^{\rm B}_{,1111}+\psi^{\rm B}_{,2222})+\psi^{\rm
             B}_{,1122} \;, \nonumber \\
  C_{\rm g}&:=\gamma_{1,11}-\gamma_{1,22}+2\gamma_{2,12}
  ={1\over 2}(\psi^{\rm E}_{,1111}+\psi^{\rm E}_{,2222})+\psi^{\rm
             E}_{,1122} \;.
             \elabel{Cfactors}
\end{align}
Thus, we see that a shear field that does not contain a B-mode
component satisfies $C_{\rm c}\equiv 0$, whereas one that has no E-mode
contribution satisfies $C_{\rm g}\equiv 0$. In the following we
provide examples for shear fields for which $C_{\rm g}\equiv 0 \equiv
C_{\rm c}$, and thus result either from an E- or a B-mode
deflection potential.

The first example is one where the deflection potential is a
polynomial of order 3. Since constant and linear terms in $\psi$ do
not cause any shear, we write
\begin{align}
\psi^{\rm E/B}&=a_{11}^{\rm E/B}\theta_1^2 + a_{12}^{\rm E/B}\theta_1 \theta_2
                +a_{22}^{\rm E/B} \theta_2^2 \nonumber \\
  &+ b_{111}^{\rm E/B} \theta_1^3
+ b_{112}^{\rm E/B} \theta_1^2\theta_2
+b_{122}^{\rm E/B} \theta_1 \theta_2^2 + b_{222}^{\rm E/B} \theta_2^3 \;.
\end{align}
This yields the linear shear field
\begin{align}
  \gamma_1&=a_{11}^{\rm E}-a_{22}^{\rm E}-a_{12}^{\rm B}
  +\rund{3 b_{111}^{\rm E}-b_{122}^{\rm E}-2 b_{112}^{\rm B}}\theta_1
            \nonumber \\
  &+\rund{b_{112}^{\rm E}-3 b_{222}^{\rm E}-2 b_{122}^{\rm B}}\theta_2\;,
  \nonumber \\
  \gamma_2&=a_{12}^{\rm E}+a_{11}^{\rm B}-a_{22}^{\rm B}
+\rund{2 b_{112}^{\rm E}+3 b_{111}^{\rm B}-b_{122}^{\rm B}}\theta_1
  \\
&+\rund{2 b_{122}^{\rm E}+b_{112}^{\rm B}-3 b_{222}^{\rm B}}\theta_2  
     \; . \nonumber       
\end{align}
It is obvious that such a linear shear field can be equally obtained
from E-mode and B-mode deflection potentials, and thus such a shear
field corresponds to an ambiguous mode. Obviously,
$C_{\rm g}\equiv 0 \equiv C_{\rm c}$ for such a field.

A less trivial example is obtained by considering axi-symmetric shear
fields of the form
\be
\gamma(\theta)=-F\rund{|\vc\theta|^2} {\theta\over \theta^*}\;,
\ee
where here we use complex notation for a vector $\vc\theta$ (i.e.,
$\theta=\theta_1+{\rm i}\theta_2$), and the asterisk denotes complex
conjugation. The term $\theta/ \theta^*={\rm e}^{2{\rm i}\vp}$, where
$\vp$ is the polar angle of $\vc\theta$, is just a phase factor. Such
a shear field is tangential to the origin at every point, and can be
generated by an axi-symmetric mass distribution $\kappa^{\rm E}$ or,
equivalently, an axi-symmetric deflection potential $\psi^{\rm
  E}$. From Eq.\,(\ref{eq:Cfactors}) we find that
$C_{\rm c}(\vc\theta)\equiv 0$ for this shear field, independent of
the function $F$. For $C_{\rm g}$, we find
\be
C_{\rm g}(\vc\theta)=-4\eck{2 F'\rund{|\vc\theta|^2}+|\vc\theta|^2
F''\rund{|\vc\theta|^2}}\;,
\ee
which is nonzero in general. However, for  $F(X)={\rm  const.}$ or
$F(X)\propto X^{-1}$, $C_{\rm g}$ also vanishes. We consider the
latter case first: it corresponds to
\be
\gamma(\theta)=-{1\over |\vc\theta|^2} {\theta\over \theta^*}
={-1\over \theta^{*2}}\;,
\elabel{PMshear}
\ee
the shear field of a point mass. Curiously, we can also get the
same shear field from a B-mode potential. Indeed, we let
\be
\psi^{\rm E}={1-f\over 2} \,\ln\rund{|\theta^2|}\;;
\quad
\psi^{\rm B}=-f\,\arctan\rund{\theta_2/\theta_1}\;,
\ee
and then we get
\be
\gamma(\vc\theta)=-{\theta_1^2-\theta_2^2+2{\rm i}\theta_1\theta_2
  \over \abs{\vc\theta}^4}\;,
\ee
in agreement with Eq.\,(\ref{eq:PMshear}), for any value of $f$. This
indeed is a curious result, stating that a pure tangential shear field
-- the ``classical'' case of an E-mode field -- can be obtained from a
B-mode potential. Putting this is different words: If we take the
tangential shear field (\ref{eq:PMshear}) and rotate the shear at
every position by 45 degree (equivalent to multiplying the shear by a
factor ${\rm i}$), then we get the classical case of a B-mode shear
field. However, this rotated field can be obtained from a pure E-mode
potential $\psi^{\rm E}=-\arctan(\theta_2/\theta_1)$. We should point
out, though, that the $\arctan(\theta_2/\theta_1)$ is not defined on
the $\theta_2$-axis where it jumps from $-\pi/2$ to $\pi/2$, and thus
the rosette-like shear field cannot be obtained from a
globally defined E-mode potential (or convergence). But if one
considers the shear field on any finite region not crossing the
$\theta_2$-axis, one cannot tell whether the shear field
(\ref{eq:PMshear}) is due to an E or a B mode.\footnote{We note that
  $\arctan(\theta_2/\theta_1)=\vp$ for $-\pi/2 < \vp< \pi/2$.  Hence,
  we could replace the $\arctan(\theta_2/\theta_1)$ just by $\vp$. In
  this case, the function would undergo only one discontinuity on a
  circle around the origin.}
  
Likewise, the shear field
\be
\gamma(\theta)=-{\theta\over \theta^*} \;,
\elabel{constshear}
\ee
which is a tangential shear field with an amplitude independent of
radius $|\theta|$, can be generated both by an E- and B-mode
deflection potential: letting
\be
\psi^{\rm E}={f-1\over 2}\, |\theta|^2 \,\ln\rund{|\theta|^2}\; $   and$
\quad
\psi^{\rm B}= f\,|\theta|^2 \, \arctan\rund{\theta_2/\theta_1}\;
\ee
leads to the shear field (\ref{eq:constshear}) for any $f$. 

\subsection{\label{sc:AppA2}Shear correlation functions from ambiguous
  shear fields}
We now consider isotropic statistical ensembles of ambiguous
shear fields and consider the resulting shear correlation
functions. For that, we consider the shear on two points on the
$\theta_1$-axis, at $\vc\theta=(\pm \vt/2,0)$, so that
$\xi_+(\vt)=\ave{\gamma(-\vt/2)\gamma^*(\vt/2)}$ and
$\xi_-(\vt) =\ave{\gamma(-\vt/2)\gamma(\vt/2)}$.\footnote{The imaginary part of these correlators vanishes due to parity
  invariance.}  Starting with the linear shear field, we consider an
ensemble of such fields, and write the shear in complex notation as
\be
\gamma(\theta)=G_2+G_1\theta+G_3\theta^*\;,
\ee
where, due to the fact that the shear is a spin-2 field, the
coefficients $G_n$ are spin-$n$ quantities that, under a rotation of
the coordinate frame, transform as
$G_n\to G_n\,{\rm e}^{-n{\rm i}\vp}$. Accordingly,
\begin{align}
  \gamma(-\vt/2)\,\gamma(\vt/2)&=G_2^2-(G_1^2+G_3^2+2G_1 G_3)\vt^2/4\;,
                                 \nonumber\\
     \gamma(-\vt/2)\,\gamma^*(\vt/2)&=\abs{G_2}^2
  -\rund{\abs{G_1}^2+\abs{G_3}^2+G_1G_3^*+G_1^*G_3} \vt^2/4    \nonumber \\  
&+\eck{G_2\rund{G_1^*+G_3^*}-G_2^*\rund{G_1+G_3}}\vt/2 \;.
\end{align}
If we now consider a statistical ensemble of such linear fields, we
have to average over the coefficients. Statistical isotropy then
implies that $\ave{G_m G_n}=0=\ave{G_m G_n^*}$ for $m\ne n$, as well
$\ave{G_nG_n}=0$, due to phase averaging over these ${\rm spin}\ne 0$
quantities. Therefore,
\begin{align}
    \xi_+(\vt)&=\ave{\gamma(-\vt/2)\,\gamma^*(\vt/2)}=
      |G_2|^2-\rund{|G_1|^2+|G_3|^2}\vt^2/4\; ,\nonumber\\
\xi_-(\vt)&=\ave{\gamma(-\vt/2)\,\gamma(\vt/2)}=0 \;,
\end{align}
which corresponds to the ambiguous modes discussed in
Sect.\,\ref{sc:AppA1}.

We next turn to the shear field caused by an ensemble of point
masses. Specifically, we consider a circular region of radius $\Theta$
in which there are $N$ point masses at locations $\vc\theta_i$ and
relative masses $m_i$, with mean mass $\ave{m}$.  At the end we
consider the limit $\Theta\to\infty$, $N\to \infty$, such that the
mean number density $\bar n=N/(\pi \Theta^2)$ is constant. The shear
field then reads
\be
\gamma(\theta)=\sum_{i=1}^N \rund{m_i\over (\theta -\theta_i)^2}^*\;.
\ee
We assume the positions $\theta_i$ of the point masses to be random
inside the circle. Therefore, the expectation value of the 
product $\gamma(-\vt/2)\gamma(\vt/2)$ is
\begin{align}
\xi_-(\vt)&=\ave{\gamma(-\vt/2)\gamma(\vt/2)}=
\eck{\prod_{n=1}^N{1\over \pi\Theta^2}\int_0^\Theta
  \d|\theta_n|\;|\theta_n|
  \int_0^{2\pi}\d\vp_n} \nonumber \\
&\times \sum_{i,j=1}^N \rund{ {m_i\over (\theta_i -\vt/2)^2}
{m_j\over (\theta_j + \vt/2)^2}}^* \;.
\label{eq:A16}
\end{align}
We now split the sum into terms $i\ne j$ and those with $i=j$. In the
former case, each term of the sum depends only on two $\theta_n$, and
the rest integrate out to unity. Those off-diagonal terms yield
\be
\sum_{i\ne j}^N {m_i m_j \over (\pi \Theta^2)^2}\, I^*(\vt/2)\, I^*(-\vt/2)\;,
\ee
where
\be
I(\vt/2)=\int_0^\Theta
  \d\theta\;\theta \int_0^{2\pi}\d\vp\;{1\over\rund{ \theta\,{\rm
        e}^{{\rm i}\vp}-\vt/2}^2} \;.
\ee  
We can now calculate the inner integral. For that, we let
$u={\rm e}^{{\rm i}\vp}$, $\d\vp=-{\rm i}\,\d u/u$, so the
$\vp$ integral becomes
\be
\int_0^{2\pi}\d\vp\;{1\over\rund{ \theta\,{\rm
        e}^{{\rm i}\vp}-\vt/2}^2}
  =-{\rm i}\oint{\d u\over u}\; {1\over ( \theta u-\vt/2)^2}\;,
  \ee
where the integral extends over the unit circle. This integral was 
calculated in \cite{Schneider96} to yield
\be
{4 \pi\over \vt^2}\eck{2 {\rm H}\rund{{\vt\over 2}-\theta}-
{\vt\over 2}\,\delta_{\rm D}\rund{\theta-{\vt\over2}}}\;.
\ee
so that $I(\vt/2)=0$ for $\Theta>\vt/2$. Thus, the off-diagonal terms
in Eq.\,(\ref{eq:A16}) do not contribute to $\xi_-$.  In fact,
$I(\theta)$ is the shear caused by a uniform disk of matter of radius
$\Theta$, and it is well known that such a disk causes no shear for
$\Theta>\theta$.

This leaves us with the diagonal terms $i=j$,
\be
\xi_-(\vt)= {N \ave{m^2}\over \pi\Theta^2}
\int_0^\Theta
  \!\!\d\theta\;\theta \oint {-{\rm i}\,\d u\over u}\;{1\over\rund{
      \theta \, u-\vt/2}^2} \;{1\over\rund{ \theta\,u+\vt/2}^2} \; .
\ee
Employing the residue theorem, we note three poles at $u_1=0$,
$u_2=\vt/(2\theta)$, and $u_3=-\vt/(2\theta)$, with
${\rm Res}(u_1)=16/\vt^4$, ${\rm Res}(u_2)={\rm
  Res}(u_3)=-8/\vt^4$. The latter two poles lie inside the unit circle
for $\theta>\vt/2$, and for this case, the contour integral
vanishes. Thus, we find
\be
\xi_-(\vt)= \bar n \ave{m^2} {4\pi\over \vt^2} \;,
\ee
corresponding to one of the ambiguous modes discussed in
Sect.\,\ref{sc:AppA1}. Repeating the calculations for the correlation
$\xi_+(\vt)$, we find that the non-diagonal terms in the double sum
vanish as well, and we are left with
\be
\xi_+(\vt)= {N \ave{m^2}\over \pi\Theta^2}
\int_0^\Theta
  \!\!\d\theta\;\theta \oint {-{\rm i}\,\d u\over u}\;{1\over\rund{
      \theta \, u-\vt/2}^2} \;{1\over\rund{ \theta /u+\vt/2}^2} \; .
\ee
The integrand in the contour integral has poles at $u_1=\vt/(2\theta)$
and $u_2=-2\theta/\vt$, and the corresponding residue are
${\rm Res}(u_1)=-16(\vt^2-4\theta^2)/(\vt^2+4\theta^2)^3$ and
${\rm Res}(u_2)=16(\vt^2-4\theta^2)/(\vt^2+4\theta^2)^3$. The former
(latter) pole is inside the unit circle for $\theta>\vt/2$
($\theta<\vt/2$). Performing the $\theta$ integral then yields
$\xi_+(\vt)=0$. 

In fact, this result could have been anticipated: the convergence
power spectrum for a random field of point masses is a constant, and
the correlation function of the convergence vanishes for any finite
separation. But the shear correlation function $\xi_+$ is identical to
the convergence correlation, so that $\xi_+(\vt)=0$ for
$\vt>0$. Furthermore, for a constant power spectrum, the second of
Eqs.\,(\ref{eq:xipmP}) shows that $\xi_-(\vt)\propto \vt^{-2}$. We
also note that the first of Eqs.\,(\ref{eq:xipmnoB}) implies that
$\xi_-(\vt)\propto \vt^{-2}$ yields $\xi_+(\vt)=0$.

We have been unable to find an analogous example of a shear field
that can be obtained from a deflection potential and which yields a
$\xi_-(\vt)\propto \vt^{-4}$ correlation. However, if we drop the
requirement that the shear field can be obtained from a potential --
for example, the shear field is due to some systematics unrelated to
the lensing effect -- then one can construct such examples. If we
consider the spin-3 field
\be
\gamma(\theta)= F\rund{|\theta|^2} \theta^3 \;,
\ee
then we find that $C_{\rm c}\equiv 0 \equiv C_{\rm g}$ if $F(X)$
satisfies the differential equation $X^2 F''+6 X F'+6F=0$. The two
independent solutions, $F\propto X^{-2}$ and $F\propto X^{-3}$, then
lead to shear fields of the form
$\gamma(\theta)\propto \theta^3/|\theta|^4$ and
$\gamma(\theta)\propto \theta^3/|\theta|^6$. Choosing the latter and constructing a random field with it, in the same way as
we did above for the point masses, we find indeed that
$\xi_-(\vt)\propto \vt^{-4}$.

\subsection{\label{sc:AppA3}Ambiguous modes in $\xi_\pm$ and their
  relation to power spectra}
We consider here the relation between shear correlation function
and the underlying power spectra, and provide examples of correlations
functions that can be derived equally well from an E- or B-mode power
spectrum, or a linear combination of both.

We start by noting that the relation between the correlation functions
and the E- and B-mode power spectra, $P_{\rm E}(\ell)$ and $P_{\rm
  B}(\ell)$, respectively,  is given by Eq.\,(\ref{eq:xipmP}).
If the correlation functions are known for all $\vt$, one can invert
these relations and get a unique decomposition into E and B modes,
\begin{align}
P_{\rm E}(\ell)&=\pi\int_0^\infty \d\vt\;\vt\,
\eck{\xi_+(\vt)\,{\rm J_0}(\ell\vt)+\xi_-(\vt)\,{\rm J_4}(\ell\vt)}
\;, \nonumber \\
P_{\rm B}(\ell)&=\pi\int_0^\infty \d\vt\;\vt\,
\eck{\xi_+(\vt)\,{\rm J_0}(\ell\vt)-\xi_-(\vt)\,{\rm J_4}(\ell\vt)} \;,
\elabel{PEBxi}
\end{align}
but on a finite interval of separations, this decomposition is not
possible.  As an example, we consider the power spectrum\footnote{We
  ignore the fact that $P_0$ can be negative; instead, we may assume
  that $P_0$ is an additive contribution to a total power spectrum
  that is positive for all $\ell$.}
\be
P_0(\ell) ={2\pi \vt_2\over \ell^2 \vt_0^2}
\eck{\ell\rund{\vt_0^2\, \xi_0+\vt_2^2\, \xi_2}{\rm J}_1(\ell\vt_2)
-2\vt_2\, \xi_2 \,{\rm J}_2(\ell \vt_2)} \;,
\elabel{ambig1}
\ee
where $\vt_2> \tmax$, and $\vt_0$ is a fiducial angular scale,
and let the E- and B-mode power spectra be $P_{\rm E}(\ell)=f
P_0(\ell)$, $P_{\rm B}(\ell)=(1-f) P_0(\ell)$. Then we find from
Eq.\,(\ref{eq:xipmP}) that 
\be
\xi_+(\vt)= \xi_0 + \xi_2\rund{\vt\over \vt_0}^2\;; \quad
\xi_-(\vt)=0
\elabel{ambig2}
\ee
for $\vt<\vt_2$, and thus for $\vt\le\tmax$, valid for any value of
$f$. Hence, we can obtain the pair of correlation functions
(\ref{eq:ambig2}) for any distribution of power on the E- and B-mode
power spectra. Therefore, we have the two ambiguous modes
$\xi_+={\rm const.}$ and $\xi_+\propto \vt^2$. We note that these
modes are ambiguous only on a finite interval. For $\vt>\vt_2$,
$\xi_+=0$, but $\xi_-\ne 0$, and in particular, $\xi_-\propto
(2f-1)$. Hence, if we had information about $\xi_\pm$ on all scales,
we could determine the parameter $f$, and the mode assignment would be
unique.

Similarly, we consider the power spectrum
\be
P_0(\ell)={2\pi \vt_0^2\over \ell^2\vt_1^3}
\eck{2\xi_{-2}\vt_1  {\rm J}_2(\ell\vt_1)
+\ell\rund{\xi_{-2}\vt_1^2+\xi_{-4}\vt_0^2}{\rm J}_3(\ell\vt_1)} \;,
\elabel{ambig3}
\ee
where $\vt_1<\tmin$. We now distribute this power as
$P_{\rm E}(\ell)=(1+f)P_0(\ell)$, $P_{\rm B}(\ell)=f P_0(\ell)$ over
E- and B modes, and then find from Eq.\,(\ref{eq:xipmP}) that
\be
\xi_+(\vt)=0\;;\quad
\xi_-(\vt)=\xi_{-2}\rund{\vt\over\vt_0}^{-2}
+\xi_{-4}\rund{\vt\over\vt_0}^{-4}\;, 
\elabel{ambig4}
\ee
which is valid for $\vt>\vt_1$ and thus for $\vt\ge \tmin$. We note that this
pair of correlation functions are independent of $f$, and thus valid
for any distribution of the power $P_0$ over E and B modes. Hence,
this is a second pair of ambiguous modes, namely $\xi_+=0$, and 
$\xi_-\propto \vt^{-2}$ and $\xi_-\propto \vt^{-4}$. Whereas
$\xi_-(\vt)=0$ for $\vt<\vt_1$, $\xi_+(\vt)\ne 0$ for smaller $\vt$,
and in particular it is proportional to $(1+2f)$. Thus, again, these modes
are ambiguous only on a finite interval.

For the more general case, we assume that the correlation
functions $\xi_+(\vt)=\xi_+^0(\vt)+\Delta\xi_+(\vt)$,
$ \xi_-(\vt)=\xi_-^0(\vt)+\Delta\xi_-(\vt)$ are written as a sum of
two terms, where the ones with a ``0'' superscript do not yield any
ambiguous modes $E^0_{a,b}=0=B^0_{a,b}$. On the other hand, we assume
that $\Delta\xi_+(\vt)$ is purely ambiguous (i.e., of the form
$ \Delta\xi_+(\vt)=\xi_0+\xi_2(\vt/\vt_0)^2$ on the finite interval
$\tmin\le\vt\le\tmax$) but has an arbitrary functional form for larger
and smaller separations. The coefficients $\xi_{0,2}$ are directly
related to the $E_{a,b}+B_{a,b}$ defined above.  From
Eq.\,(\ref{eq:PEBxi}), we then find
\be
P_0:=\Delta P_{\rm E}(\ell)+\Delta P_{\rm B}(\ell)
=2\pi \int_0^\infty \d\vt\;\vt\,{\rm J}_0(\ell\vt)\,\Delta\xi_+(\vt)\;.
\ee
We again distribute the power over modes in the form
$\Delta P_{\rm E}(\ell)=f P_0(\ell)$,
$\Delta P_{\rm B}(\ell)=(1-f)P_0(\ell)$, and then calculate
$\Delta\xi_-(\vt)$ on the finite interval,
\begin{align}
\Delta\xi_-(\vt)
&=(2f-1)\int_0^\infty\d\ell\;\ell\,{\rm J}_4(\ell\vt)
\int_0^\infty\d\theta\;\theta\,{\rm
  J}_0(\ell\theta)\,\Delta\xi_+(\theta) \nonumber\\ 
&= (2f-1)\Bigg\{\Delta\xi_+(\vt)
+\int_0^\tmin\!\d\theta\;\theta\,
\Delta\xi_+(\theta)\rund{{4\over \vt^2}-{12\theta^2\over \vt^4}}\nonumber\\ 
&\quad +
\int_\tmin^\vt\d\theta\;\theta\,\eck{\xi_0+\xi_2\rund{\vt\over\vt_0}^2}\rund{{4\over
    \vt^2}-{12\theta^2\over \vt^4}} \Bigg\} \\
&= (2f-1)\Bigg\{{1\over \vt^2}
\eck{4\int_0^\tmin\!\!\!\!\d\theta\;\theta\,\Delta\xi_+(\vt)
-2\vt_{\rm min}^2\xi_0-{\vt_{\rm min}^4\over\vt_0}\,\xi_2}\nonumber\\
&+{1\over \vt^4}
\eck{12\int_0^\tmin\!\d\theta\;\theta^3\,\Delta\xi_+(\vt)
+3\vt_{\rm min}^4\xi_0+{2\vt_{\rm min}^6\over\vt_0^2}\xi_2}\Bigg\} \;,\nonumber
\end{align}
where we made use of the relation
\[
\int_0^\infty\!\!\!\d\ell\;\ell\,
{\rm J}_0(\ell\vt)\,{\rm J}_4(\ell\theta)
={1\over\vt}\delta_{\rm D}(\vt-\theta) 
+\rund{{4\over\theta^2}-{12\vt^2\over \theta^4}}
{\rm H}(\theta-\vt) \;, 
\]
where $\delta_{\rm D}$ and ${\rm H}$ denote the Dirac delta `` function''
and the Heaviside step function, respectively. We see that
$\Delta\xi_-(\vt)$ only contains ambiguous modes inside the finite
interval and that their amplitudes depend on the integral of
$\Delta\xi_+$ over scales below $\tmin$; in other words, the amplitudes are assumed to be
unmeasured. Because of this, the fraction $f$ of B-mode power
attributed to the $\Delta\xi_\pm$ cannot be determined. We can go
through the analogous exercise to fix $\Delta\xi_-$ and calculate
$\Delta\xi_+$, which then only contains ambiguous modes with an
amplitude that depends on $f$ and moments of $\xi_-$ taken over scales
larger than $\tmax$.

\section{\llabel{newCOSEBI}A new set of COSEBIs}
The $\mu=n$ coefficients in Eq.\,(\ref{eq:EBdef}) define the
COSEBIs. These COSEBIs depend on the choice of the weight functions
$T_{\pm n}(\vt)$. We point out that the $T_{\pm n}(\vt)$ used in this
paper differ from those in SEK in their dimensions: whereas in SEK,
these filter functions were chosen to be dimensionless, we chose them
here to have dimension $\rm (angle)^{-2}$, as can be seen from
Eq.\,(\ref{eq:ON-rel}). Correspondingly, the COSEBIs defined here are
dimensionless, whereas they have dimension $\rm (angle)^{2}$ in
SEK. We think the current choice is more natural than the earlier one.

Furthermore, our orthonormality relation (\ref{eq:ON-rel}) differs
from that of SEK through the factor $\vt$ in the integral. This new
definition allowed us to show that the $T_{-n}(\vt)$ also form an
orthonormal basis, which they do not with the orthonormality relation
used in SEK.

In SEK, we constructed two sets of functions $T_{\pm n}(\vt)$, one polynomials in $\vt$ and the other polynomials in $\ln(\vt)$,
termed linear and logarithmic COSEBIs, respectively. The latter were
shown to be more convenient, in that fewer COSEBI modes are needed to
extract the full cosmological information contained in mode-separable
correlation functions.

\subsection{Linear COSEBIs}
We consider the case of polynomial COSEBIs first, for which we
transform the interval $\tmin\le\vt\le\tmax$ onto the interval
$-1\le x\le 1$ via
\be
\vt=\bar\vt (1+Bx) \;.
\elabel{xtheta}
\ee
We then set $T_{\pm n}(\vt)=\bar\vt^{-2} t_{\pm n}(x)$. Since
$\d\vt=B\bar\vt\,\d x$, we then see from Eqs.\,(\ref{eq:ON-rel}) and
(\ref{eq:Tminusortho}) that the $t_{\pm n}(x)$ obey the orthonormality
relations 
\be
\int_{-1}^1 \d x\;(1+Bx)\,t_{\pm m}(x)\,t_{\pm n}(x) =\delta_{mn} \;.
\elabel{ON-t}
\ee
This equation also motivates the pre-factor in the orthonormality
relation (\ref{eq:ON-rel}). Furthermore, the constraints
(\ref{eq:conditions}) are translated into
\be
\int_{-1}^1\d x\;(1+Bx)\,t_{+n}(x) = 0
=\int_{-1}^1\d x\;(1+Bx)^3\,t_{+n}(x) \;.
\elabel{conditions3}
\ee
The functions $T_{-n}(\vt)=\bar\vt^{-2}t_{-n}(x)$ are then calculated
from 
\begin{align}
t_{-n}(x)=t_{+n}(x)&+{4B\over (1+Bx)^2}
\int_{-1}^x \d y\; (1+By)\,t_{+n}(y)  \nonumber \\
&-{12B\over (1+Bx)^4}\int_{-1}^x \d y\; (1+By)^3\,t_{+n}(y) \;.
\elabel{tminus}
\end{align}
We constructed a set of polynomial functions $t_{+n}(x)$ obeying the
orthonormality relation (\ref{eq:ON-t}) and the constraints
(\ref{eq:conditions3}), where $t_{+n}(x)$ is a polynomial of
$(n+1)$-th order, given by
\begin{align}
t_{+1}(x)&={(5-B^2)\,P_2(x)-3 B \,P_1(x)+B^2
           \,P_0(x)\over\sqrt{10-6B^2}} \;,
\nonumber\\
t_{+n}(x)&=\eck{2 (n+2)\, B\, P_{n+1}\rund{1\over B}\,
  P_{n+2}\rund{1\over B}}^{-1/2} 
\elabel{tpns}\\
&\times
\sum_{k=0}^{n+1}(-1)^{k} (2k+1)\, P_k\rund{1\over B}\, P_k(x)\quad 
{\rm for}\quad n\ge 2 \;.
\nonumber
\end{align}
The sign of the $t_{+n}(x)$ has been chosen such that $t_{+n}(-1)>0$,
implying $T_{+n}(\tmin)>0$.  In order to show the validity of this
result, we first consider, for $n\ge 2$, the expression
\begin{align}
(1+Bx)& \sum_{k=0}^{n+1}(-1)^{k} (2k+1) 
P_k\rund{1\over B} P_k(x)\nonumber\\
&=
B\sum_{k=0}^{n+1}(-1)^{k}\Bigg\{{2k+1\over B} P_k\rund{1\over B}
  P_k(x) \\
&+P_k\rund{1\over B}\eck{(k+1)P_{k+1}(x)+k P_{k-1}(x)}\Bigg\} \;,\nonumber
\end{align}
where we used the recursion relation for Legendre polynomials,
$P_{k}$. Changing the summation index for the last two terms as
$k\to k\pm 1$ and applying the recursion relation for Legendre
polynomials again, this time for the $P_k(1/B)$, we see that only two
terms survive, and we obtain
\begin{align}
(1+&Bx) \sum_{k=0}^{n+1}(-1)^{k} (2k+1) P_k\rund{1\over B} P_k(x) 
 \\
&=
   (-1)^{n+1} (n+2) B \eck{P_{n+1}\rund{1\over B}P_{n+2}(x)+P_{n+2}
   \rund{1\over B}P_{n+1}(x)} \;.\nonumber
\end{align}
Therefore, we find that, for $n\ge 2$,
\begin{align}
(1+Bx)\,t_{+n}(x)
&=(-1)^{n+1}\sqrt{(n+2)B\over 2P_{n+1}(1/B) P_{n+2}(1/B)} \nonumber\\
\times&\eck{P_{n+1}\rund{1\over B}P_{n+2}(x)
+P_{n+2}\rund{1\over B}P_{n+1}(x)} \;.
\end{align}
Using the orthogonality relation of the Legendre polynomials, it is
then straightforward to show that the orthonormality relation
(\ref{eq:ON-t}) is satisfied for $m,n\ge 2$. Furthermore, since
$(1+Bx)t_{+n}$ for $n\ge 2$ contains no term $P_k(x)$ with $k\le 2$,
the orthogonality relation is clearly valid for $m=1$, $n\ge
2$. Finally, it is easy to see that conditions
(\ref{eq:conditions3}) are satisfied for $n\ge 2$, and for $n=1$, it
can be shown from straightforward integration.  Hence, the system
(\ref{eq:tpns}) forms the set of polynomial weight functions we were
looking for.

It should be stressed that these functions are easy to calculate: for
a given survey setup, one needs to calculate the $P_k(1/B)$ only
once, and the $P_n(x)$ are easily obtainable from the recursion
relation of the Legendre polynomials. Whereas it is possible in
principle to obtain explicit expressions for the corresponding
functions $t_{-n}(x)$, this may not be needed: since the calculation
of COSEBIs requires the calculation of the $\xi_\pm$ and the $T_{-n}$
at a large number of $\vt$-values \citep[see][]{Asgari15}, it is probably
computationally more efficient to evaluate the integrals in
Eq.\,(\ref{eq:tminus}) using very small increments in the upper bound $x$.

\begin{figure*}
\small
\begin{verbatim} 
Nmax=20; tmin=1; tmax=400; tbar=(tmax+tmin)/2; BB=2(tmax-tmin)/tbar; zm=Log[Rationalize[tmax/tmin]]
gamm[a_,z_]=Gamma[a,0,z]
Do[J[k,j]=Re[N[gamm[j+1,-k zm]/(-k)^(j+1),130]],{k,2,4},{j,0,2 Nmax+1}]
Do[Do[a[n,j]=J[2,j]/J[2,n+1]; a[n+1,j]=J[4,j]/J[4,n+1],{j,0,n}]; b[n]=-1; b[n+1]=-1;
   Do[a[m,j]=NSum[J[2,i+j] c[m,i],{i,0,m+1}, WorkingPrecision->80, NSumTerms->Nmax],{m,1,n-1},{j,0,n}];
   Do[bb[m]=-NSum[J[2,i+n+1] c[m,i],{i,0,m+1}, WorkingPrecision->80, NSumTerms->Nmax],{m,1,n-1}];
   Do[a[m,j]=a[m,j]/bb[m],{m,1,n-1},{j,0,n}]; Do[b[m]=1,{m,1,n-1}];
   A=Table[a[i,j],{i,1,n+1},{j,0,n}]; B=Table[b[i],{i,1,n+1}];
   CC=LinearSolve[A,B]; Do[c[n,j]=CC[[j+1]],{j,0,n}]; c[n,n+1]=1;
   tt[n,z_]=Simplify[Sum[c[n,j] z^j,{j,0,n+1}]];
   roots=NSolve[tt[n,z]==0,z];Do[r[n,j]=roots[[j,1,2]],{j,1,n+1}];
   t[n,z_]=Product[(z-r[n,j]),{j,1,n+1}];
   normgral=NIntegrate[Exp[2z] t[n,z]^2,{z,0,zm},WorkingPrecision->50];
   norm[n]=Sqrt[tbar^2 BB/tmin^2/normgral]; t[n,z_]=t[n,z] norm[n],   {n,1,Nmax}]
ROOTS=Table[N[r[n,j],8],{n,1,Nmax},{j,1,Nmax+1}]; NORM=Table[N[norm[n],8],{n,1,Nmax}]
\end{verbatim}
\caption{Mathematica \citep{WolfMathe} program to calculate the roots in
  Eq.\,(\ref{eq:tlogfns}). They are stored with eight significant
  digits in the lower-left half of the table {\tt ROOTS}, and
the table {\tt NORM} contains the normalization coefficients, $N_n$}
\flabel{matheprog}
\end{figure*}

\subsection{Logarithmic COSEBIs}
The roots of the polynomial weight functions $T_{+n}(\vt)$ are fairly
uniformly distributed over the interval $\tmin< \vt<\tmax$. The shear
correlation functions $\xi_\pm(\vt)$ vary more strongly for smaller
$\vt$ than for larger $\vt$, and therefore are expected to contain
more (cosmological) information on these smaller scales. Therefore, it
is useful to consider a set of weight functions $T_{+n}(\vt)$ that
also show more structure on smaller scales, as done before in SEK. We
let
\be
T_{+n}(\vt)={1\over \bar\vt^2}\,t_{+n}\rund{\ln{\vt\over\tmin}}\;,
\ee
so that the functions $t_{+n}(z)$ are defined for
$0\le z\le \ln(\tmax/\tmin)=z_{\rm m}$. The constraints
(\ref{eq:conditions}) and the orthonormality relation
(\ref{eq:ON-rel}) then read in terms of the $t_{+n}$:
\begin{align}
\int_0^{z_{\rm m}}\d z\;{\rm e}^{2z}\,t_{+n}(z)&=0 \;,\nonumber \\
\int_0^{z_{\rm m}}\d z\;{\rm e}^{4z}\,t_{+n}(z)&=0 \; ,\elabel{teqs}\\
\int_0^{z_{\rm m}}\d z\;{\rm e}^{2z}\,t_{+n}(z)\,t_{+m}(z)&=
{B\bar\vt^2\over\vt_{\rm min}^2}\,\delta_{mn} \nonumber \;.
\end{align}
We now choose the $t_{+n}(z)$ to be polynomials of order $n+1$, and
write them in the form
\be
t_{+n}(z)=\sum_{k=0}^{n+1}c_{nk}\,z^k\;. 
\ee
The equations (\ref{eq:teqs}) then lead to a linear system of equations
for the coefficients $c_{nk}$, as was shown in SEK. Indeed, this
system is very similar to the corresponding one in SEK, and differs
only in the definition of the orthonormality relation for the
$T_{+n}$. Hence, we refer the reader to SEK for details of the method
how the solution for the $c_{nk}$ is obtained. As was mentioned there,
one needs the $c_{nk}$ to have very high numerical precision, in
particular for large values of $\tmax/\tmin$. However, if we write the 
polynomials in the form
\be
t_{+n}(z)=N_n \prod_{i=1}^{n+1} (z-r_{ni}) \;,
\elabel{tlogfns}
\ee
then a moderate precision for the roots $r_{ni}$ is sufficient. As an
example, for $\tmax/\tmin=400$ and eight significant digits of the
$r_{ni}$, the orthonormality relations for the first 20 $T_{+n}$ are
satisfied to better than $10^{-18}$. In Fig.\,\ref{fig:matheprog}, we
display a Mathematica \citep{WolfMathe} program that calculates the
roots $r_{ni}$.

An expression for the  corresponding function
$T_{-n}(\vt)=t_{-n}[\ln(\vt/\tmin)]/\bar\vt^2$ can then be calculated
from Eq.\,(\ref{eq:Tplusminus}), yielding
\be
t_{-n}(z)=t_{+n}(z)+\int_0^z\d y\; t_{+n}(y)\eck{4 {\rm e}^{2(y-z)}
  -12{\rm e}^{4(y-z)}}\;.
\ee
Hence, the $t_{-n}$ can be easily calculated as numerical integrals
over the $t_{+n}$ in the form (\ref{eq:tlogfns}).

\begin{figure}
  \includegraphics[width=\columnwidth]{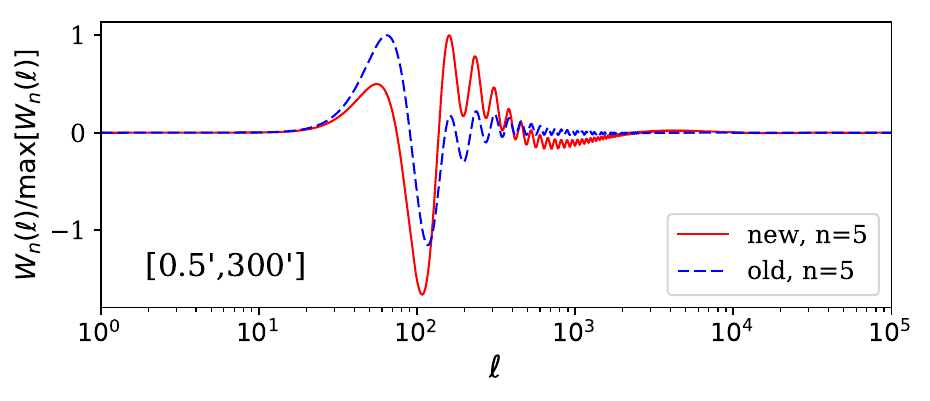}
  \caption{Comparison between the new dimensionless (solid red) and
    the old SEK (dashed blue) COSEBIs. We show the form of the fifth
    COSEBI weight function, $W_5(\ell)$. Each curve is normalized
    with respect to its maximum value.  We chose an angular
    separation interval of $0.5$ to $300$ arcminutes to define the
    weights.}
  \label{fig:W5}
\end{figure}

The COSEBIs are related to the underlying power spectrum by the integral
\be
E_n=\int_0^\infty {\d\ell\;\ell\over 2\pi} P_{\rm E}(\ell)\,W_n(\ell)\;,
\ee
where the weight function $W_n$ is given by
\be
W_n(\ell)=\int_\tmin^\tmax \d\vt\;\vt\,T_{+n}(\vt)\,{\rm J}_0(\vt\ell)\;.
\ee
These weight functions thus describe the sensitivity of the COSEBIs to
the power spectrum. As an example, we plot in Fig.\,\ref{fig:W5} the
function $W_5(\ell)$ and compare it to the corresponding one of the
COSEBIs defined in SEK, in both cases for the logarithmic weight
functions. As can be seen, the ``new'' $W_5$ is significant nonzero
over a somewhat broader range in $\ell$. It is this feature that makes
the new COSEBIs less correlated than the old ones, as shown in
Fig.\,\ref{fig:corr}. On the other hand, the wider $\ell$-range may
lead to an increase in the sensitivity of the COSEBIs to different
baryonic feedback effects, compared to that of the SEK COSEBIs
\citep[see][]{Asgari20}, which shall be explored in future work.

In particular, it must be stressed that the information content of the
SEK COSEBIs and the dimensionless COSEBIs are exactly the same, if
their full (infinite) sets are considered; in fact, one can transform
one set into the other. The difference in the properties illustrated
in Figs.\,\ref{fig:corr} and \ref{fig:W5} are not due to the different
orthonormality relations, but due to the specific choice of polynomial
weight functions $T_{+n}(\vt)$. Different sets of weight functions may
be constructed, for example to make the first $N$ of the $W_n(\ell)$ more
localized and thus potentially less sensitive to baryonic effects.

\section{\llabel{subCOSEBI} COSEBIs on a subinterval}
\def\tminp{{\vt^\prime_{\rm min}}}
\def\tmaxp{{\vt^\prime_{\rm max}}}
In this section we consider the relation between the COSEBIs on a
subinterval $\tminp\le\vt\le\tmaxp$, and the original ones on
$[\tmin,\tmax]$, where $\tmin\le\tminp < \tmaxp\le\tmax$. We denote
with $B'$ and $\bar\vt'$ the relative width and the mean angle inside
the subinterval. Furthermore, we denote by $T'_{\pm\mu}(\vt)$ the basis
functions on the subinterval, which have a support on this
subinterval. The coefficients $\tau'_{\pm\mu}$, 
defined in analogy with Eqs.\,(\ref{eq:EBplus}) and
(\ref{eq:EBminus}), are then obtained from the correlation functions
$\xi_\pm$ by
\be
\tau'_{\pm\mu}=\int_{\tminp}^{\tmaxp}\d\vt\;\vt\,T'_{\pm\mu}(\vt)\,
\xi_\pm(\vt) = \sum_\nu {\cal T}^\pm_{\mu\nu}\,\tau_{\pm\nu}\;,
\ee
where we used representation (\ref{eq:xiexp}) of the correlation
function and defined
\be
{\cal T}^\pm_{\mu\nu}={\bar\vt^2\over B}
\int_{\tminp}^{\tmaxp}\d\vt\;\vt\,T'_{\pm\mu}(\vt)\,T_{\pm\nu}(\vt)\;.
\elabel{calTdef}
\ee
Using the relation between the $\tau_{\pm n}$ and the COSEBIs $E_n$,
$B_n$, we obtain
\begin{align}
E_\mu'&={\tau'_{+\mu}+\tau'_{-\mu}\over 2}={1\over 2}\sum_\nu
\eck{\rund{{\cal T}^+_{\mu\nu}+{\cal T}^-_{\mu\nu}}E_\nu
+\rund{{\cal T}^+_{\mu\nu}-{\cal T}^-_{\mu\nu}}B_\nu } \;,\nonumber \\
B_\mu'&={\tau'_{+\mu}-\tau'_{-\mu}\over 2}={1\over 2}\sum_\nu
\eck{\rund{{\cal T}^+_{\mu\nu}-{\cal T}^-_{\mu\nu}}E_\nu
+\rund{{\cal T}^+_{\mu\nu}+{\cal T}^-_{\mu\nu}}B_\nu } \;.\nonumber
\end{align}
We now look at some properties of the transfer matrices ${\cal
  T}^\pm$. Since the functions $T'_{-m}$ and $T_{-n}$ are related to 
$T'_{+m}$ and $T_{+n}$ though the transformation (\ref{eq:transff}),
we can apply the Lemma in Sect.\,2 and obtain from
Eq.\,(\ref{eq:calTdef}) that
\be
{\cal T}^-_{mn}={\cal T}^+_{mn} \;.
\elabel{calTpm}
\ee
Furthermore, for $\nu=a,b$, the functions $T_{+\nu}(\vt)$ are of the
form $x_0+x_2 \vt^2$. From the analog of conditions
(\ref{eq:conditions}) for the $T'_{+\mu}$ functions, we then infer
that
\be
{\cal T}^+_{ma}=0={\cal T}^+_{mb} \;.
\ee
Similarly, for $\nu=a,b$, the functions $T_{-\nu}(\vt)$ are of the
form $x_2 \vt^{-2}+x_4\vt^{-4}$, so that the condition
(\ref{eq:conditions2}) yields
\be
{\cal T}^-_{ma}=0={\cal T}^-_{mb} \;.
\ee
Together, we than find that
\be
E'_m=\sum_{n=1}^\infty {\cal T}^+_{mn}\,E_n \;;
\quad
B'_m=\sum_{n=1}^\infty {\cal T}^+_{mn}\,B_n \;.
\ee
This result then shows that the E- and B-mode COSEBIs on the
subinterval can be calculated from the E- and B-mode COSEBIs on the
original angular interval. The transfer matrix ${\cal T}^+$ depends on
the choice of basis functions; in general we expect that in order to
obtain $E'_m$ to a given accuracy, one needs to use $E_n$'s up to
significantly larger $n$. However, subdividing the angular interval
into subintervals, as has been done in some previous work, does not
yield any additional information if one chooses the maximum order of
COSEBIs properly.

Since in general, ${\cal T}^\pm_{an}$ and ${\cal T}^\pm_{bn}$ will be
nonzero, the ambiguous modes in the subinterval will not only depend
on the ambiguous modes on the full interval, but some E and B modes of
the full interval will be transferred to the ambiguous modes on the
subinterval. This is to be expected: the smaller the angular range
is, the more pure-mode information gets lost to the ambiguous modes. 
\end{appendix}

\end{document}